\documentclass[10pt]{article}
\usepackage{appendix}
\usepackage{algorithm}
\usepackage{algpseudocode}
\usepackage{graphicx}
\usepackage{float}
\usepackage{geometry}
\usepackage{wrapfig}
\usepackage{pdfpages}
\usepackage{booktabs}
\usepackage{amsmath}
\usepackage{placeins}
\usepackage{amssymb}
\usepackage{amsthm}

\usepackage{graphicx}

\usepackage{cite}
\usepackage{color} 
\usepackage[colorlinks,breaklinks]{hyperref}
\hypersetup{hypertexnames=true,
  linkcolor=blue,anchorcolor=black,citecolor=blue,urlcolor=blue
}

\DeclareGraphicsExtensions{.png,.pdf}
\graphicspath{{imglocal/}{img/}{./}} 

\usepackage[labelfont=bf,labelsep=period,justification=raggedright]{caption}
\usepackage{subcaption}

\makeatletter
\renewcommand{\@biblabel}[1]{\quad#1.}
\makeatother

\newcommand{\giacomo}[1]{}

\date{}

\begin{document}

\begin{flushleft}
{\Large
\textbf{An event-based architecture for solving constraint satisfaction problems}
}
\\
Hesham Mostafa*, 
Lorenz K. M\"uller*, 
Giacomo Indiveri.
\\
Institute for Neuroinformatics, University of Zurich and ETH Zurich, Switzerland \\
E-mail: \{hesham,lorenz,giacomo\}@ini.uzh.ch\\ 
*These authors contributed equally to this work
\end{flushleft}
\begin{abstract}
Constraint satisfaction problems (CSPs) are typically solved using conventional von Neumann computing architectures. However, these architectures do not reflect the distributed nature of many of these problems and are thus ill-suited to solving them. In this paper we present a hybrid analog/digital hardware architecture specifically designed to solve such problems. We cast CSPs as networks of stereotyped multi-stable oscillatory elements that communicate using digital pulses, or events. The oscillatory elements are implemented using analog non-stochastic circuits. The non-repeating phase relations among the oscillatory elements drive the exploration of the solution space. We show that this hardware architecture can yield state-of-the-art performance on a number of CSPs under reasonable assumptions on the implementation. We present measurements from a prototype electronic chip to demonstrate that a physical implementation of the proposed architecture is robust to practical non-idealities and to validate the theory proposed. 
\end{abstract}


Constraint satisfaction problems (CSPs) are a fundamental class of problems in computer science with wide applicability in areas such as channel coding~\cite{MacKay03}, circuit optimization~\cite{kirkpatrick_etal83}, and scheduling~\cite{garey_etal76}. Algorithms for solving CSPs are typically run on von Neumann architectures where a number of processing units compute using a shared memory pool. These architectures were not explicitly developed to solve CSPs. This raises the question: How can we construct a more efficient computing substrate whose architecture better reflects the distributed nature of CSPs? In this paper we address this question by describing a distributed dynamical system whose dynamics execute an efficient search for CSP solutions and which can be easily implemented using Complementary Metal-Oxide Semiconductor (CMOS) Very Large Scale Integration (VLSI) electronic chips.

Many dynamical systems 
 that have been proposed for solving CSPs violate the `physical implementability' condition~\cite{Zhang_Constantinides92,Nagamatu_Yanaru96,Ercsey-Ravasz_Toroczkai11}. Non-physicality arises from the use of variables that can grow without bounds as the system is searching for solutions. On the other hand, there is a long well-established tradition of developing physically realizable dynamical systems, e.g., in the form of artificial neural networks, to solve CSPs or ``Best-match problems''~\cite{Minsky_Papert69,Rumelhart_McClelland86}. Early attempts in this field used attractor networks, such as Hopfield networks~\cite{Hopfield82}, to solve CSPs like the traveling salesman problem~\cite{Hopfield_Tank85,Hopfield_Tank86}. These attractor networks, however, would often get stuck at locally optimal problem solutions (so called ``local minima''). 
To avoid getting stuck at local minima,  stochastic attractor networks were proposed~\cite{Habenschuss_etal13,Maass14} which make use of explicit sources of noise to force the network to continuously explore the solution space.
While noise is an inextricable part of any physical system, dynamically controlling the noise power to balance ``exploratory'' versus ``greedy'' search, or  to realize an annealing schedule that moves the network from an exploratory phase to a greedy one, is not a trivial  operation and puts an additional overhead on the physical implementation. 

The architecture we propose in this paper makes use of analog oscillator circuits that, once fabricated, inevitably have incommensurable frequencies (i.e., frequencies that are not rational multiples of each other). Rather than requiring external sources of noise or relying on random fluctuations, our architecture exploits the non-repeating phase relations among the analog oscillators to drive the search for optimal solutions. This greatly reduces the requirements for the design of the hardware implementation, as it does not require precise circuits. Indeed, as the system exploits, by construction,  the inhomogeneities  present in physical devices, it is possible to design extremely compact and simple oscillator circuits, that are allowed to have large mismatch figures. Each variable in a CSP is represented by a node consisting of an analog oscillator and a state-holding asynchronous digital circuit. To achieve robust and scalable computation, the nodes communicate using digital pulses, or events. This combination of analog and digital circuits running in a hybrid continuous/event-driven mode avoids many of the problems that affect pure analog VLSI systems such as susceptibility to noise, degradation of analog signals during storage and communication, and signal restoration/refresh issues.

This architecture is inspired by computational neuroscience studies that used oscillatory rate-based neural networks to solve constraint satisfaction problems~\cite{Mostafa_etal13b}. The architecture we present here uses different dynamics that were developed primarily based on engineering considerations to be as robust and easily implementable as possible. The architecture was also developed to be general enough to allow the instantiation of various efficient algorithms for solving CSPs. We present results from a CMOS implementation of this architecture on a prototype VLSI chip and show that the chip operation reproduces the dynamics of the simulated system.
Our results expose a surprising relation between the dynamics of coupled multi-stable oscillators and the search for CSP solutions and highlight a novel mode of distributed, parallel, mixed analog/digital computation that can form the basis of various hardware/physical systems for solving CSPs.



\section*{Results}
\subsection*{Description of the architecture}

\begin{figure}[h]
 \centering
     \begin{subfigure}[b]{0.4\textwidth}
     \includegraphics[width=\textwidth]{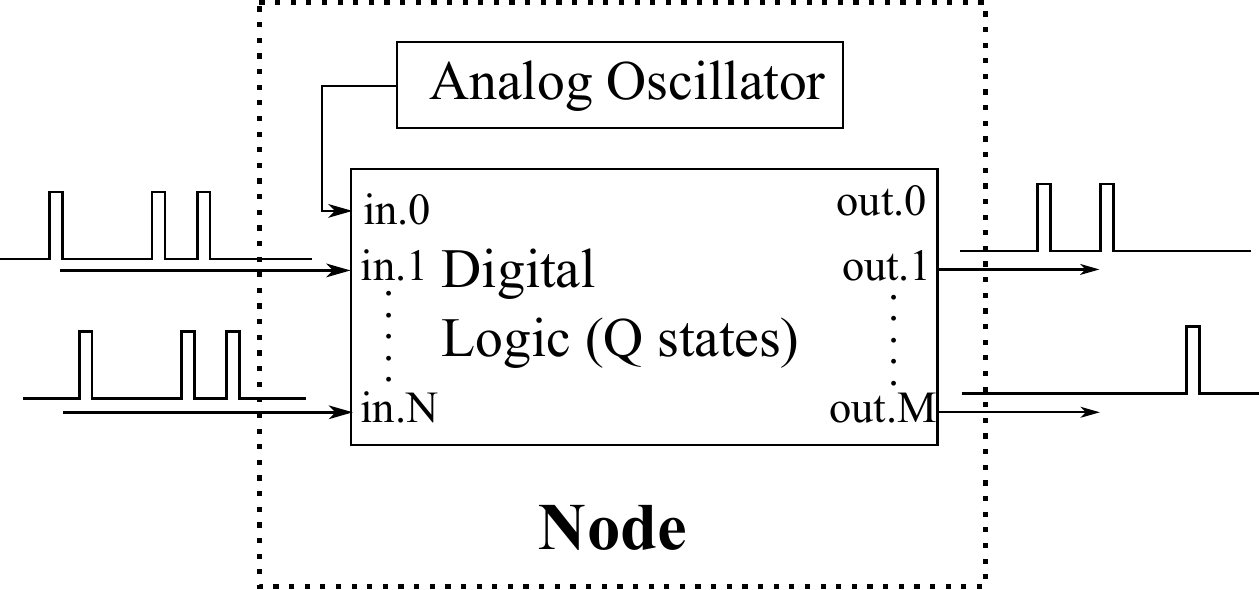} 
     \subcaption{}
     \label{fig:variable_a}
   \end{subfigure}
   \quad\quad\quad
   \begin{subfigure}[b]{0.4\textwidth}
     \begin{minipage}{2.3in}
     Event: on receiving an event on $in.i$\\
     \hspace*{0.2in} $r \leftarrow g(i,s)$ \\
     \hspace*{0.2in} $s \leftarrow f(i,s)$\\
     \hspace*{0.2in} generate event on out.r
     \end{minipage}
     \subcaption{}
     \label{fig:variable_b}
   \end{subfigure}
   \\
   \begin{subfigure}[b]{0.6\textwidth}
     \begin{minipage}{2.3in}
       \begin{equation*}
         f(i,s) =
         \begin{cases}
           s & \text{if}\; i=0 \\
           i       & \text{otherwise} 
         \end{cases}
         \quad\quad
         g(i,s) =
         \begin{cases}
           s & \text{if}\; i=0 \\
           0       & \text{otherwise} 
         \end{cases}
         \label{eq:toy}
       \end{equation*}
     \end{minipage}
     \subcaption{}
     \label{fig:variable_c}
     \end{subfigure} \\
     \begin{subfigure}[b]{0.6\textwidth}
       
       \includegraphics[width=\textwidth]{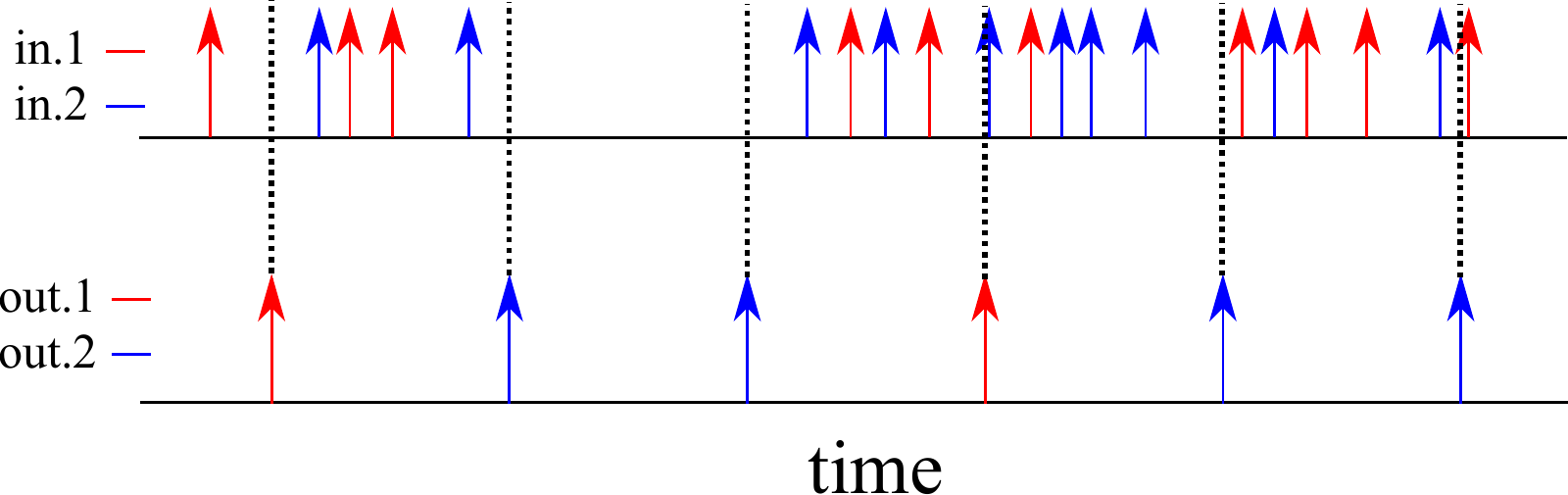} 
       \subcaption{}
       \label{fig:variable_d}
     \end{subfigure}

   \caption{(\subref{fig:variable_a}) General form of a node composed of an analog oscillator and digital logic. The digital logic is event-driven. (\subref{fig:variable_b}) On an input event, the digital logic evaluates the index of the output port, updates its state, and generates an output event in that order according to the functions $f$ and $g$. (\subref{fig:variable_c}) Definition of the $f$ and $g$ functions for an example binary node. (\subref{fig:variable_d}) Simulation of the example node showing how its output events reflect the identity of the last input event it received.}
\label{fig:variable}
\end{figure}

The proposed event-based architecture for solving CSPs is composed of a network of nodes which communicate via digital events. A node is shown schematically in Fig.~\ref{fig:variable_a}. Each node has $N$ externally accessible input ports, one internal input port, $M$ output ports, and one dummy output port. The analog oscillator in the node generates a continuous stream of digital events which are sent to the node's internal port: $in.0$. The digital logic in the node has an internal state $s$ which can take one of $Q$ possible values. On the arrival of an event on any of the input ports, the node's digital logic evaluates the index of the output port to which it should send the event based on the index of the triggered input port and on the current state of the digital logic; it updates its internal state; and it transmits the event via the output port selected (see Fig.~\ref{fig:variable_b}). Selection of the `dummy' output port $out.0$ is equivalent to suppressing the event. The digital logic is fully described by the event routing function $g$ and the state update function $f$ which are both deterministic. Given their analog nature, the frequencies of the oscillators in the different nodes are not rational multiples of each other. This requirement is trivial to achieve in a VLSI design of analog oscillators. 

For solving CSPs, a subset of the nodes in the network will represent the actual problem variables while others will represent helper variables that encode other problem-relevant quantities (for example, whether a constraint is satisfied or not). The value of a variable/node at a point in time is the index of the output port on which the node emitted its last event. Thus, a variable/node with $M$ output ports can have $M$ possible values. The output port of one node can connect to the input ports of one or more nodes and one input port can receive events from multiple output ports. One output port can {\it not} be connected to multiple input ports on the same node. In the following sections, we describe how to connect nodes/variables together and how to define the nodes/variables behavior so as to solve a number of hard CSPs. The procedure to map a CSP to this distributed architecture depends on the type of the CSP but in general, the mapping is done so that the distributed and parallel dynamics of the network of nodes tries to put the problem variables/nodes in a state where their outputs satisfy all the constraints


Figures~\ref{fig:variable_c} and~\ref{fig:variable_d} show the definition and illustrate the behavior of an example node which has two input ports, two output ports, and two possible internal states ($N=M=Q=2$). The state of the example node/variable is the index of the last external event it received and the node/variable advertises its state by generating an event on the corresponding output port when it receives an event on the internal port $in.0$ as shown in Fig.~\ref{fig:variable_c} (we refer to this as `updating'). Assume this example node is the target node receiving events from multiple sources nodes. Since it is only the last received event that determines the value advertised by an updating node, the phase relations between the analog oscillators in the network determine which of the source nodes generates the decisive event that determines the event generated by the target node. This would be the source node that updated just before the target node updates. The phase relations are continuously changing in an aperiodic manner since the oscillation frequencies are incommensurable. The shifting phase relations thus continuously change which source node manages to influence the output events of the target node.

For the node described in Fig.~\ref{fig:variable_c}, assume $N_1$ nodes are trying to force a target node to $1$ and $N_2$ nodes are trying to force it to $2$, the fraction of $1$ events generated by the target node roughly goes to $\frac{N_1}{N_1+N_2}$ (assuming the difference in oscillator frequencies of the source nodes are small) if observed for a long enough time. Thus, the more nodes that try to force a target node to a particular value, the more likely the target node is to output that value, yet there is always a chance that even a single source node that is in conflict with the majority will update just before the target node updates, thereby causing the target node to go against the majority influence. As we will show in the next sections, this behavior can be exploited to allow the network to escape from local minima where flipping a single variable/node may increase the number of violated constraints. However, a node will {\it never} take a value that is in conflict with all incoming influences which is why the globally optimal state is stable. We show that this mostly greedy, but sometimes exploratory, behavior can be exploited to efficiently solve a variety of hard CSPs.

\subsection*{Boolean Satisfiability Problems}
\label{sec:bool}
Let ${\bf X} = \{x_1,\ldots,x_N\}$ be a set of boolean variables. A literal is either a variable or its negation. The solution to a boolean satisfiability or K-SAT problem is the variable assignment that satisfies the logical expression involving the variables of ${\bf X}$:
$$c_1 \wedge c_2 \wedge \cdots \wedge c_m = \text{True}$$
where the clause $c_i$ is the disjunction of $K$ literals. n-SAT for $n \geq 3$ is NP-complete \cite{Sipser96}.




\subsubsection*{The probSAT Algorithm}
One of the most efficient algorithms for solving SAT problems is the probSAT algorithm~\cite{Balint_Schoning12}, which iteratively modifies a variable assignment by choosing a random unfulfilled clause $c_u$ and changing the assignment of (`flipping') a random variable $x_f$ in $c_u$, thereby fulfilling $c_u$. The choice $x_f$ is governed by a heuristic function $f(m,b)$, where $m$ (the `make' heuristic) is the number of clauses that are newly fulfilled when $x_f$ is flipped and $b$ (the `break' heuristic) is the number of clauses that are newly unfulfilled when $x_f$ is flipped. The heuristic function is renormalized into a probability over the available choices and $x_f$ is chosen according to these probabilities.
The heuristic function $f(m,b)$ can take several different forms. In our benchmarks, we use the particularly effective `exponential' form:
\begin{equation}
f(m,b) = \frac{x^m}{y^b}
\end{equation}
where $x$ and $y$ are parameters. 

\subsubsection*{Mapping Break-only probSAT to a network of nodes}
\begin{figure}
  \centering
     \includegraphics[width=0.5\textwidth]{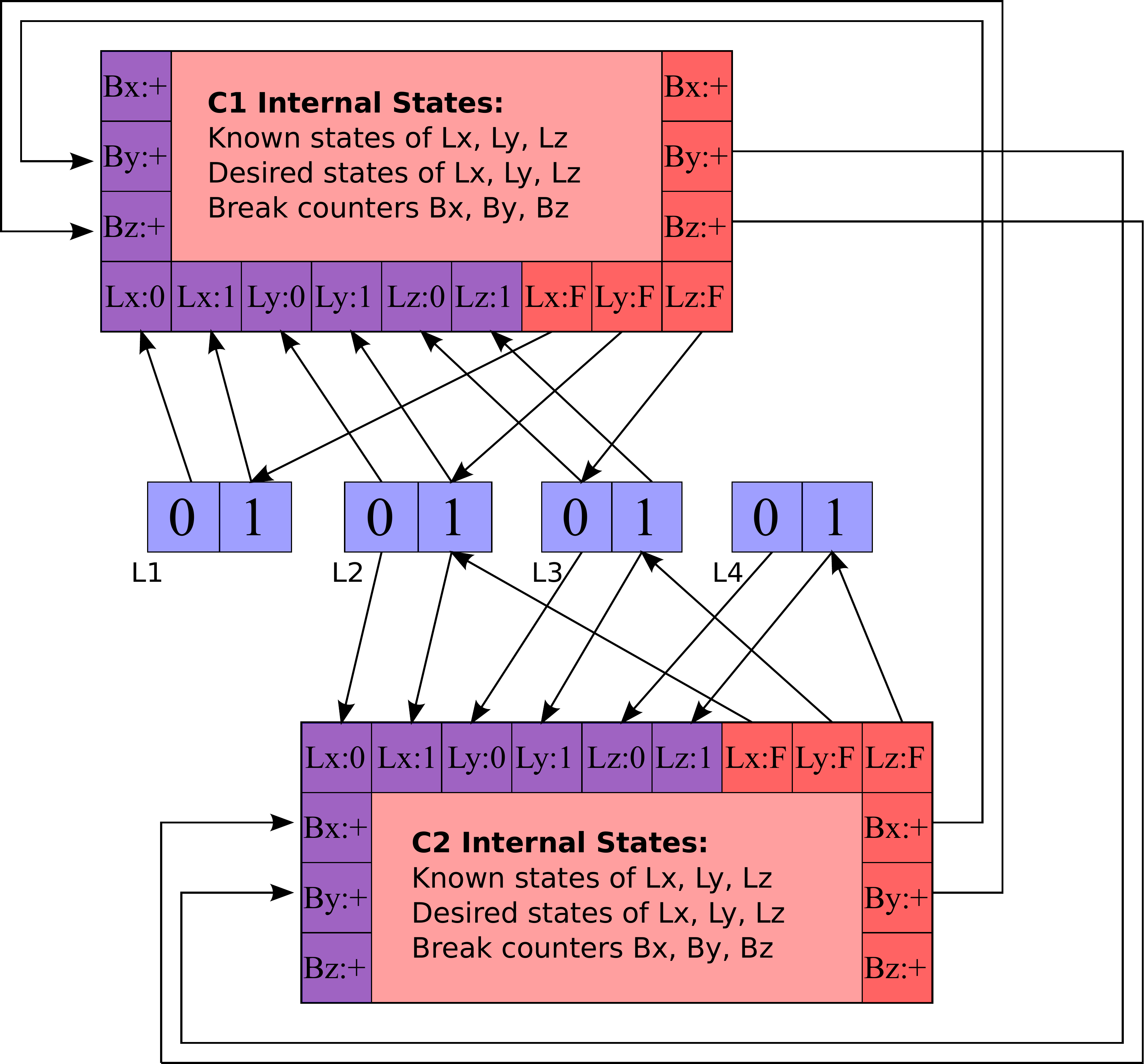} 
     \caption{Network corresponding to the example SAT problem $C1 \wedge C2$ where $C1 = (L1 \vee L2 \vee \neg L3)$ and $C2 = (L2 \vee L3 \vee L4)$. For the constraints $C1$ and $C2$, the squares at the edge of the box indicate input ports (purple) and output ports (red). Events are routed along the arrows. Input/output behavior of both types of nodes is described in the main text.}
\label{fig:net_spa}
\end{figure}

The probSAT algorithm that only uses the `break' heuristic can be loosely mapped to our architecture by using two types of nodes: nodes representing variables and nodes representing constraints/clauses (see Fig.~\ref{fig:net_spa}).
We consider only 3-SAT problems but extensions to n-SAT for $n > 3$ are straightforward. Each variable node has two states. It updates and advertises its state (by generating an event on one of its two output ports) whenever it receives an event from a clause node. Additionally, it advertises its state whenever it receives an event from the internal oscillator on input port $in.0$. 

When the clause node receives a break event (event arriving on one of the three break input ports, one corresponding to each variable), it increments the corresponding break counter. On events from the internal oscillator, a clause node evaluates what state the connected variables have last advertised. If there is no variable in a `fulfilling' state, the clause node sends an  event to flip the variable with the smallest associated `break' count  and sends a `break' event to every constraint this variable is connected to in order to indicate that the flipped variable is the only variable keeping the constraint fulfilled. If there is exactly one `fulfilling' variable, the clause node sends a `break' event to every constraint that the variable is connected to. If there is more than one `fulfilling' variable, the clause node does not send out any events. The break counters are reset after each event from the internal oscillator.  
An unfulfilled clause node thus always chooses to flip the variable with minimal break count (with ties resolved according to a fixed variable ordering). This flip heuristic is deterministic and simpler than the heuristic employed by standard probSAT.

\subsubsection*{Network and Sequential probSAT: Performance Comparison}
We compare the performance of the network to that of the standard (sequential) probSAT algorithm \cite{Balint_Schoning12}. We evaluate the network performance in two cases: The `ideal' case where events are transmitted instantly and never lost; and a `non-ideal' case where events have a delay uniformly distributed between zero and ten percent of the node oscillation period and a ten percent chance to get lost completely. The non-ideal case simulates the imperfections of an actual physical implementation where spike delivery is neither instantaneous nor guaranteed.

\begin{figure}
 \centering
     \begin{subfigure}[h]{0.4\textwidth}
     \includegraphics[width=\textwidth]{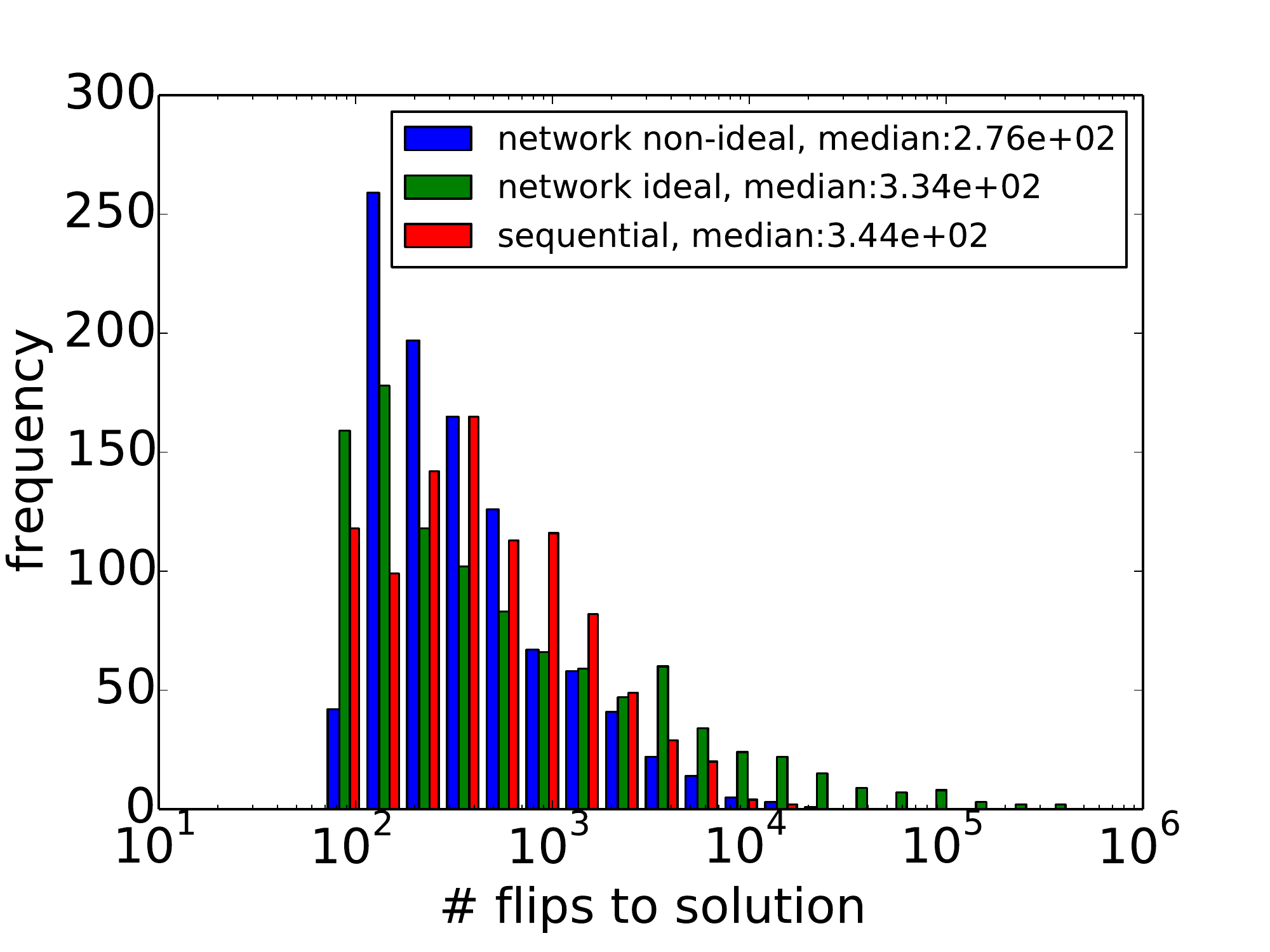} 
     \subcaption{exponential, flips}
     \label{fig:net_noErr_exp_flip}
   \end{subfigure}
   \quad
   \begin{subfigure}[h]{0.4\textwidth}
     \includegraphics[width=\textwidth]{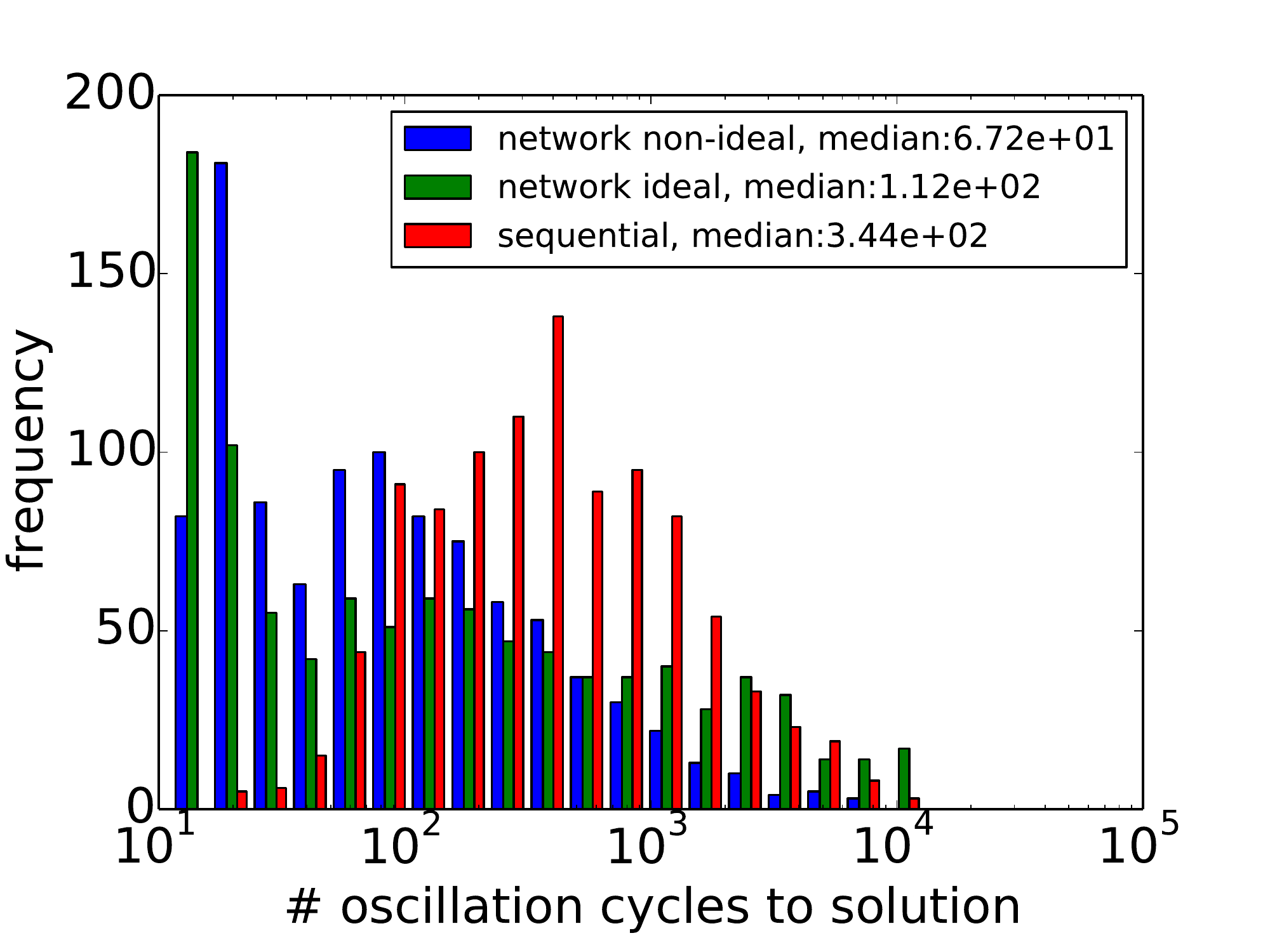} 
     \subcaption{exponential, cycles}
     \label{fig:net_noErr_exp_cyc}
   \end{subfigure}
   \caption{Performance comparison of network and standard probSAT with exponential heuristic function ($x=1$, $y=2.06$). Note that the histograms have logarithmically spaced and sized bins. The best performing algorithm in both metrics is the network with lost and delayed events. In plot~\ref{fig:net_noErr_exp_cyc}, the distribution looks bimodal because for very small numbers of cycles to solution, this number could not be accurately measured (network convergence is checked every 20 cycles).}
\label{fig:net_noErr}
\end{figure}

The first benchmark is a set of 1000 intermediate size, difficult 3-SAT problems taken from SATLIB \cite{Hoos_Stutzle98} with 50 variables and 218 clauses each. Figure~\ref{fig:net_noErr_exp_flip} shows the number of variable flips needed to reach the solution for the ideal and non-ideal network and for standard probSAT. The median flips to solutions is smallest for the non-ideal network. 
This indicates that using incommensurable oscillators to effectively ``randomize'' clause update order is better than choosing a clause to fulfill at random. 
Surprisingly, the non-ideal network performs better than the ideal one. Losing/delaying events might increase network efficiency by making it more exploratory as clauses now have imperfect information about the state of the variable nodes.

Figure~\ref{fig:net_noErr_exp_cyc} shows the average number of oscillation cycles (per variable) needed to find the solution. For standard probSAT, we assume one cycle corresponds to one variable flip. This measure takes into account the fact that the network, due to its distributed nature, can update variables in parallel. The number of oscillation cycles indicates how fast a hardware implementation of the network would need to be to run faster than the standard algorithm on a conventional CPU (which achieves around 1-10 Mega-Flip per second). Since the average number of oscillation cycles to solution is about a fifth of the number of flips performed by the sequential algorithm, a hardware implementation would need to use analog oscillators with an average frequency of 0.2-2 MHz to perform as well as a conventional computer. 

We can not give a comparison with state-of-the-art benchmarks because our software simulator is only able to simulate $10$ cycles for each node per second if used to simulate networks implementing large modern benchmarks. Based on the solution times of standard probSAT, an estimated $10^6-10^9$ cycles are needed to solve a single current day problem (i.e. 1-1000 days to simulate the corresponding network).

\begin{figure}
 \centering
     \begin{subfigure}[h]{0.3\textwidth}
     \includegraphics[width=\textwidth]{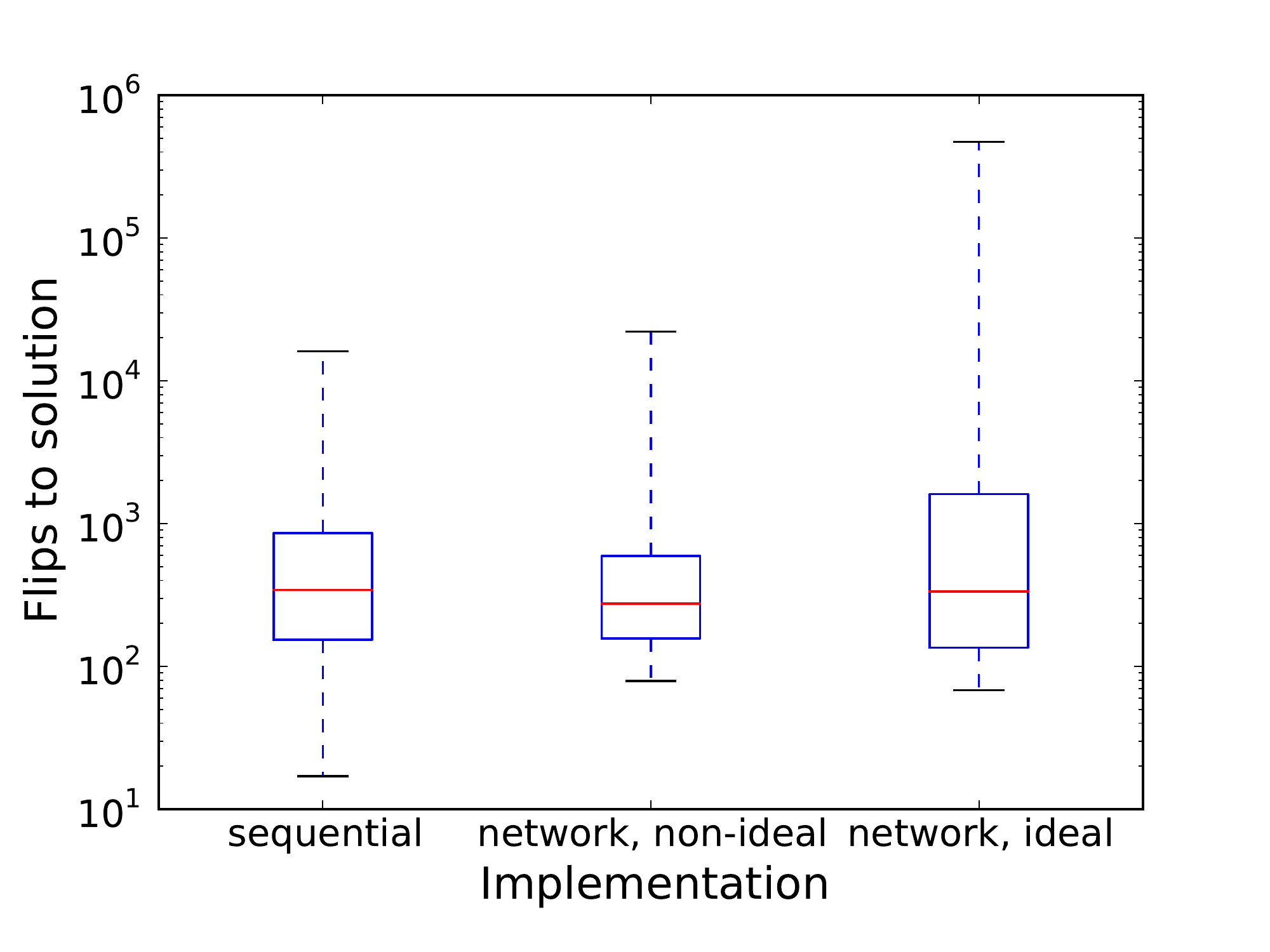} 
     \subcaption{50 literals, 218 constraints}
     \label{fig:scale_50}
   \end{subfigure}
   \quad
   \begin{subfigure}[h]{0.3\textwidth}
     \includegraphics[width=\textwidth]{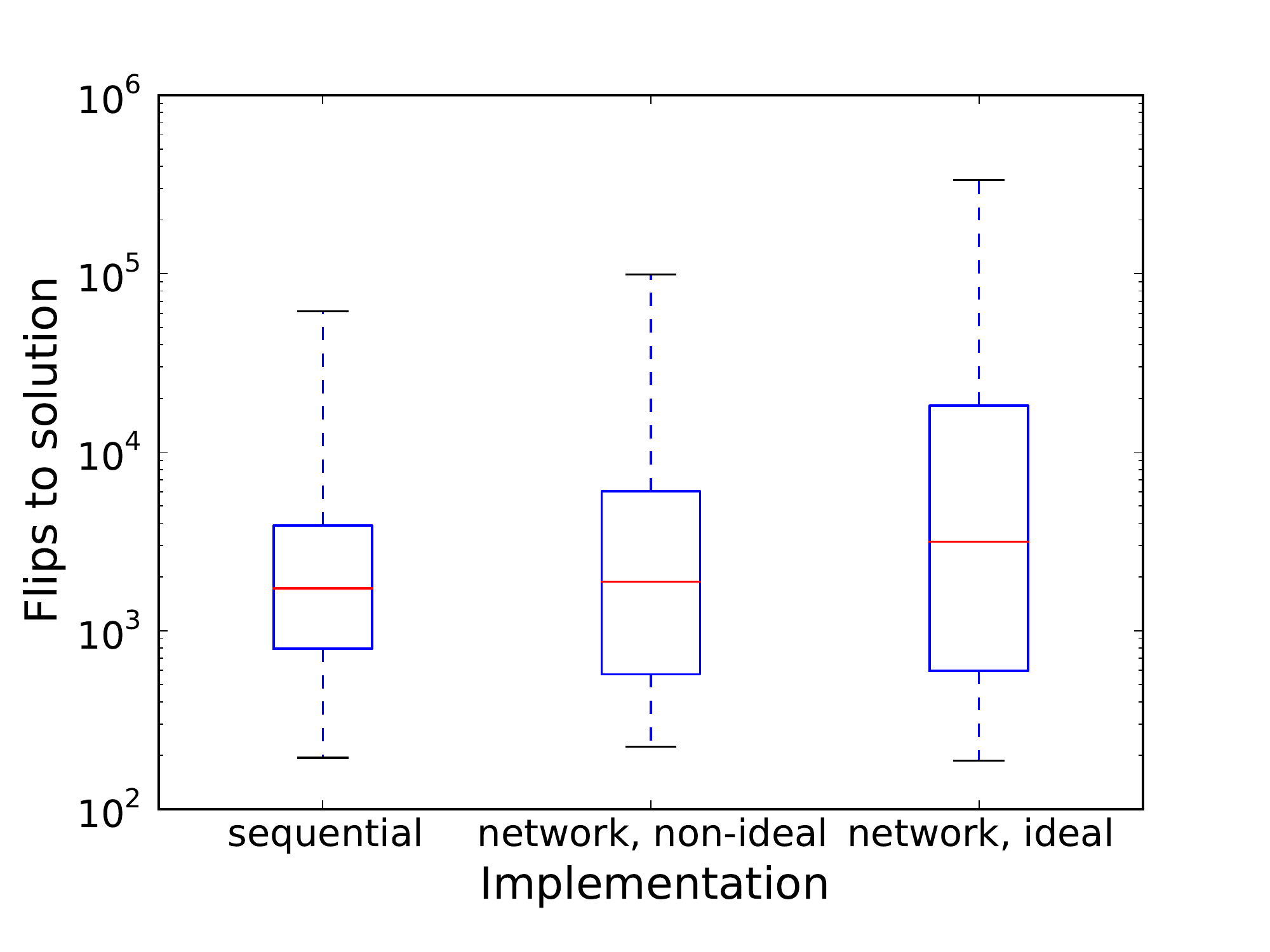} 
     \subcaption{100 literals, 430 constraints}
     \label{fig:scale_100}
   \end{subfigure}
   \quad
    \begin{subfigure}[h]{0.3\textwidth}
     \includegraphics[width=\textwidth]{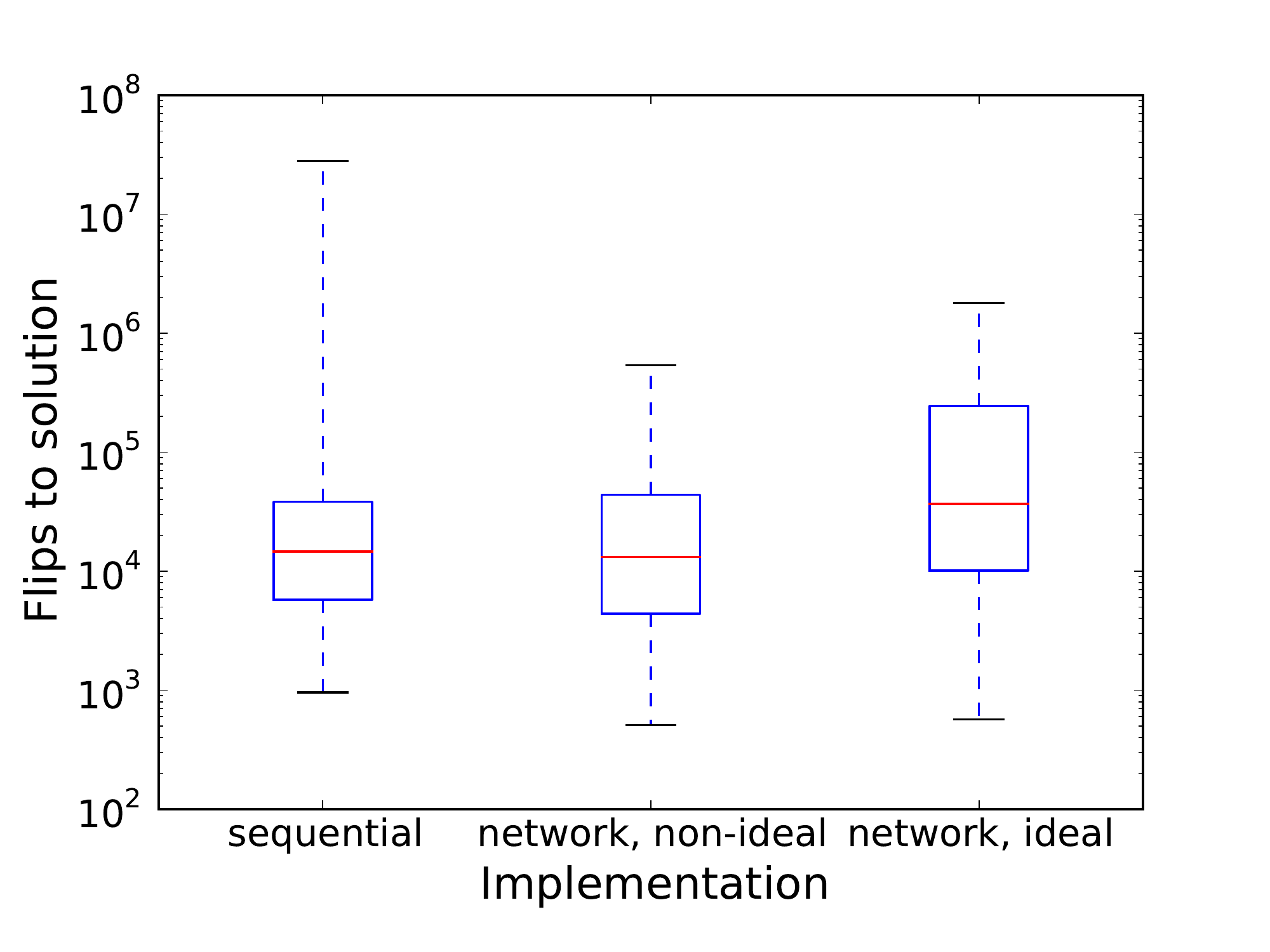} 
     \subcaption{200 literals, 860 constraints}
     \label{fig:scale_200}
   \end{subfigure}
   \caption{Performance comparison of network and standard probSAT with exponential heuristic function ($x=1$, $y=2.06$) on different problem sizes with $100$ trials per problem.  The red line indicates the median, the box outlines the 1st and 3rd quartile and the whiskers show the full range of the data. In plot~\ref{fig:scale_200}, the ideal network did not converge in 8 of 100 cases within the given $1.5\cdot 10^8$ cycles. Sequential and network formulations scale similarly well with problem size.}
\label{fig:scale}
\end{figure}

As an alternative, we evaluate the performance on various problem sizes to ensure that the network performance scales equally well as standard probSAT; since this is the case (see Fig.~\ref{fig:scale}), it is reasonable to assume that for large problems, the network performs as well as the standard algorithm in terms of number of flips to solution. 

\subsection*{Graph coloring problem}
\label{sec:colour}
A $k$-coloring for a graph $G$ with vertices $V(G)$ and a set of edges $E(G)$ is a map $\phi: V(G) \rightarrow \{1,2,\ldots,k\}$. In the graph coloring problem, the goal is to find a proper k-coloring $\phi_0$ of $G$ where $\phi_0(x) \neq \phi_0(y)$ for all $\lbrace x,y \rbrace \in E(G)$.

\subsubsection*{Network implementation and performance}
\label{sec:kalg}
\begin{figure}
  \centering
     \includegraphics[width=0.5\textwidth]{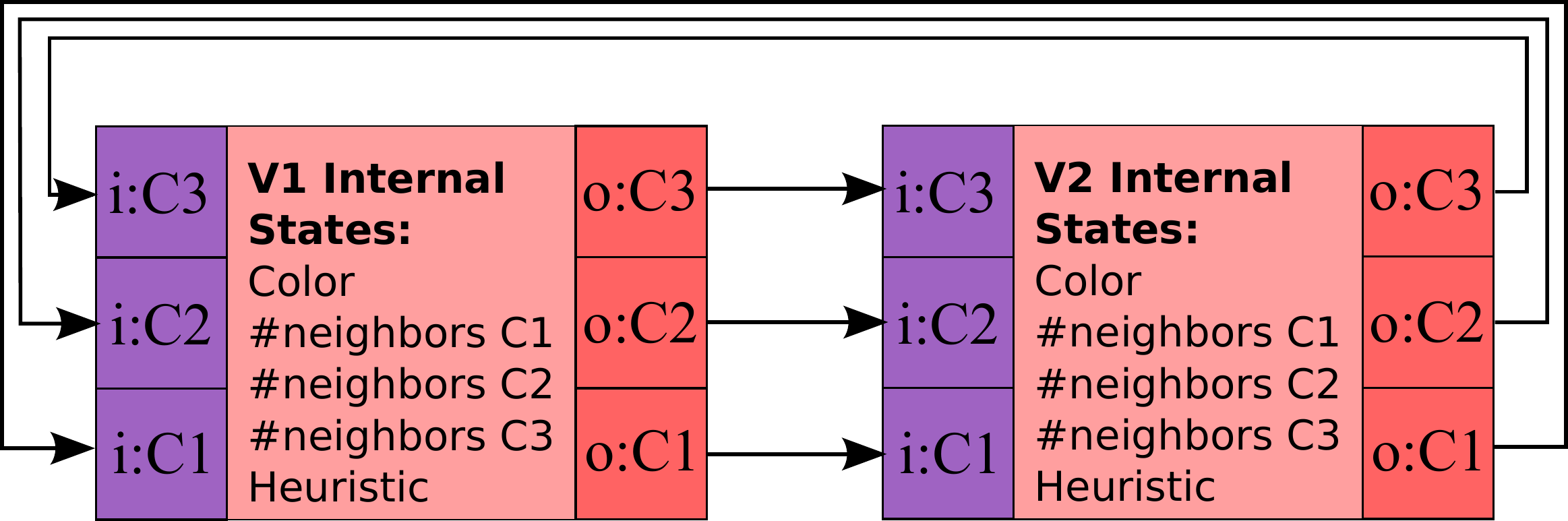} 
     \caption{Network corresponding to the 3-coloring of the graph $V=\lbrace V1,V2\rbrace$, $E=\lbrace (V1,V2)\rbrace$. The squares at the edge of the box indicate input ports (purple) and output ports (red). Events are routed along the arrows.}
\label{fig:net_kcolor}
\end{figure}

To solve a k-coloring problem, we map each vertex in the graph to a network node with $k$ input ports and $k$ output ports as shown in Fig.~\ref{fig:net_kcolor}. Whenever the internal oscillator in a node/vertex generates an event, the node advertises its color by generating an event on one of the $k$ output ports. Events from a node/vertex are routed to all its neighbors in the graph. Each node maintains k counters that count how many of its neighbors have a particular color. These counters are incremented when a node receives events from its neighbors. At an internal oscillator event, if the counter corresponding to the current node color is non-zero (one of the neighbors has the same color), the node chooses a different color. If the internal boolean variable, `heuristic', is true, the node chooses the color with the fewest conflicts (smallest neighbor count). If `heuristic' is false, the node chooses the next color in a fixed arbitrary ordering of colors. The node then resets the $k$ counters, flips the `heuristic' binary variable and generates an event to advertise its color. A min conflict heuristic thus takes turns with a heuristic free scheme to update a conflicting node in each cycle.


We assessed the performance of this algorithm on several k-coloring problems of intermediate difficulty (see Table~\ref{tab:kcol}) taken from \cite{Ruiz_Romay11} in which a different massively parallel coloring algorithm (`gravitational swarm intelligence' (GSI)) was assessed. As in the previous section on boolean satisfiability, we cannot attempt state-of-the-art sized problems since the software simulation of a large network takes an infeasibly long time. In terms of average numbers of oscillation cycles to solution the network compares favorably to GSI \cite{Ruiz_Romay11}.

\begin{table}
\centering
\begin{tabular}{l r r c r r r }
\toprule
Graph & \#vertices & \#edges & Density & K & \#GSI & \#network \\
\midrule
myciel7 & 191 & 2360 & 0.13 & 8 & 302 & \textbf{145} \\
myciel6 & 95 & 755 & 0.17 & 7 & 92 & \textbf{31} \\
myciel5 & 47 & 236 & 0.21 & 6 &  97 & \textbf{19} \\
myciel4 & 23 & 71 & 0.28 & 5 & 25 & \textbf{3} \\
myciel3 & 11 & 20 & 0.36 & 4 &   21 & \textbf{2} \\
david & 87 & 986 & 0.21 & 11 & 208 & \textbf{95} \\
anna & 138 & 812 & 0.21 & 11 & 300 & \textbf{8} \\
huck &74 & 662 & 0.22 & 11 & 84 & \textbf{8} \\
jean & 80 & 508 & 0.16 & 10 & 165 & \textbf{16} \\
queen 5x5 & 25 & 160 & 0.53 & 5 & \textbf{302} & N/A \\ 
1\_fullins\_3 & 30 & 100 & 0.23 & 4 & 37 & \textbf{11} \\
1\_fullins\_4 & 93 & 593 & 0.14 & 5 & \textbf{76} & 366 \\
1\_fullins\_5 & 282 & 3247 & 0.08 & 6 & \textbf{222} & 1593 \\
2\_fullins\_3 & 52 & 201 & 0.15 & 5 & 67 & \textbf{47} \\
2\_fullins\_4 & 212 & 1621 & 0.07 & 6 & 176 & \textbf{120} \\
miles\_250 & 128 & 387 & 0.04 & 8 & \textbf{317} & 2021 \\
\bottomrule
\end{tabular}
\caption{Number of cycles to convergence on common k-coloring benchmarks~\cite{Ruiz_Romay11} of our network and a massively parallel algorithm~\cite{Ruiz_Romay11}. Each number in the network column is an average of 4 runs with redrawn oscillator frequencies; one run for the queens graph did not converge in $10^5$ cycles (the other runs averaged $530$ steps to convergence).}
\label{tab:kcol}
\end{table}

\subsection*{Implementing analog costs and the traveling salesman problem}
The problems considered so far, 3-SAT and graph coloring problems, have hard constraints that should all be satisfied. Our architecture can also handle CSPs with weighted constraints by exploiting the relations between the frequencies of the analog oscillators. We illustrate this approach using the traveling salesman problem (TSP). A TSP with N cities is defined by the $N \times N$ matrix ${\bf D}$ where $d_{ij}$ is the distance from city $i$ to city $j$. The goal is to find a minimum length closed tour of all the cities. We can not require a distributed architecture like ours to stabilize at the optimal solution as there is no general way, short of using long-term memory resources for storing previously visited tours, to verify that a tour is optimal. Our goal is thus to map a TSP to a network of nodes that continuously explores all possible valid tours but that has a higher chance of visiting tours with smaller distances. 

\begin{figure}[h]
 \centering
 \begin{subfigure}[b]{0.3\textwidth}
     \includegraphics[width=\textwidth]{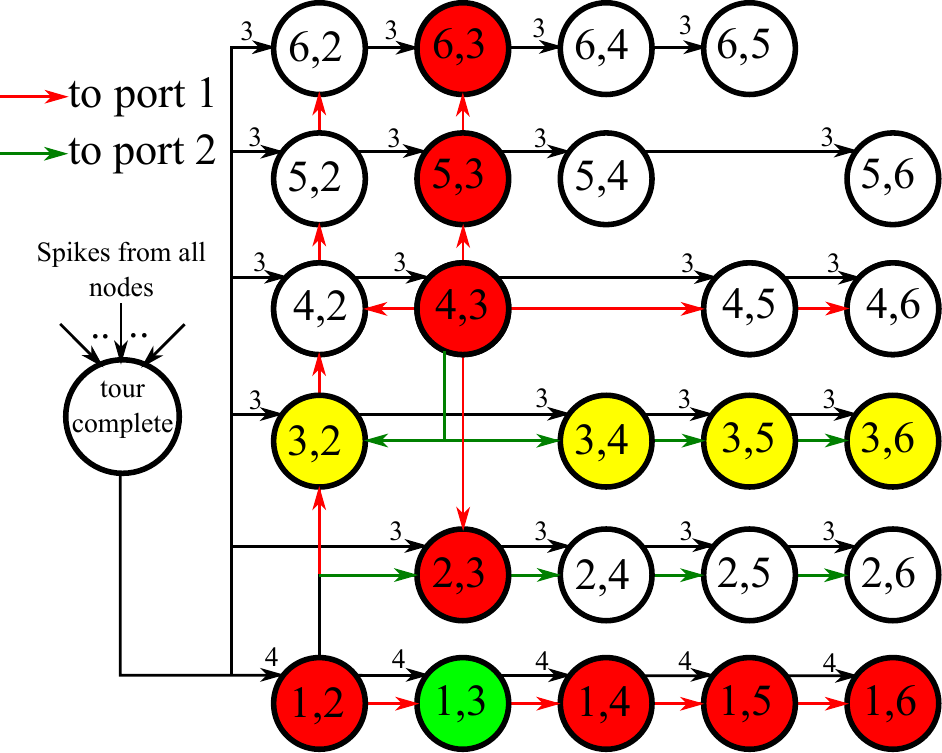} 
     \subcaption{}
     \label{fig:tsp_a}
   \end{subfigure}
   \quad
   \begin{subfigure}[b]{0.3\textwidth}
     \includegraphics[width=\textwidth]{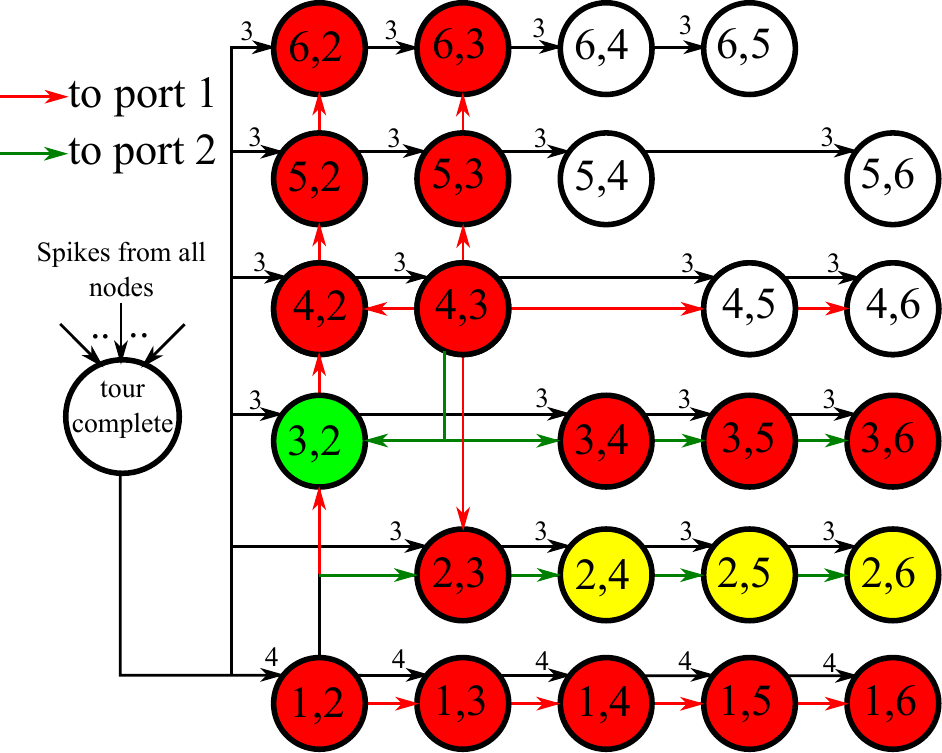} 
     \subcaption{}
     \label{fig:tsp_b}
   \end{subfigure}
   \quad
   \begin{subfigure}[b]{0.3\textwidth}
     \includegraphics[width=\textwidth]{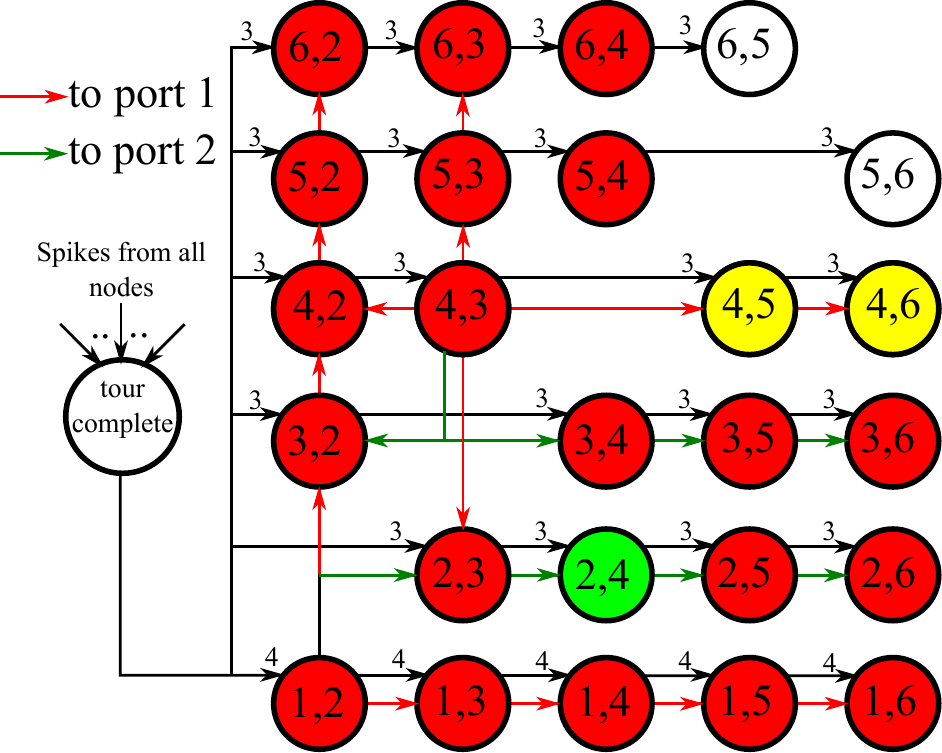} 
     \subcaption{}
     \label{fig:tsp_c}
   \end{subfigure} \\

   \begin{subfigure}[b]{0.5\textwidth}
     \includegraphics[width=\textwidth]{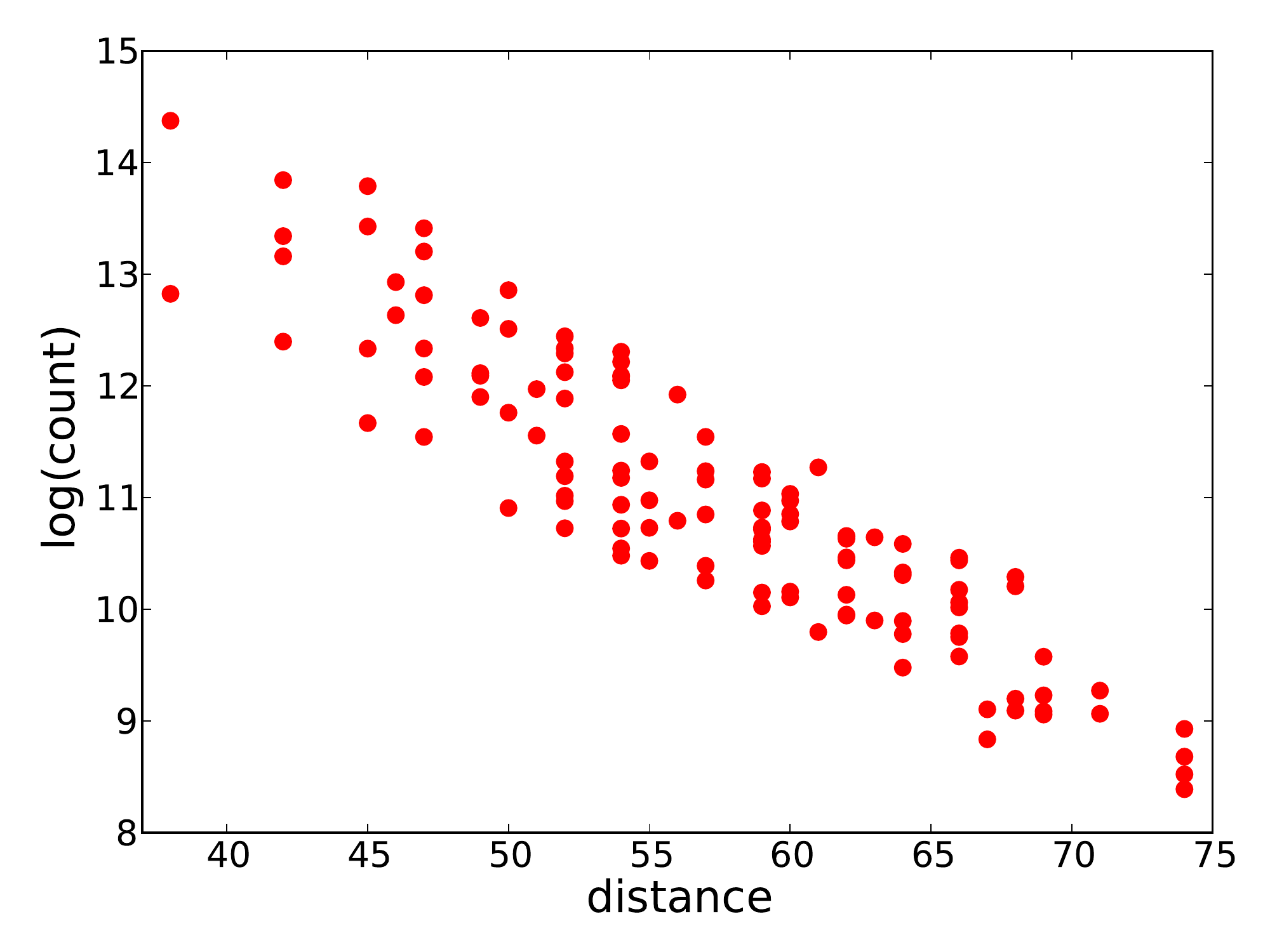} 
     \subcaption{}
     \label{fig:tsp_d}
   \end{subfigure}

   \caption{(\subref{fig:tsp_a}, \subref{fig:tsp_b}, \subref{fig:tsp_c}) TSP with 6 cities. Red and green arrows indicate inputs to ports $1$ and $2$ respectively. Numbers next to black arrows denote target input ports. Only the routing of the events of edge nodes $(1,2)$ and $(4,3)$ and of the `tour completion' node are shown. A green node is a node that has just generated an event, red nodes are in state 1 (inactive), yellow nodes are in state 3 (active). \subref{fig:tsp_a}, \subref{fig:tsp_b}, \subref{fig:tsp_c} show the network state after each event for 3 successive events.  (\subref{fig:tsp_d}) Frequency of occurrence of each tour in $1.5*10^7$ tours generated by a network implementing a six cities TSP problem as a function of the tour distance.}
\label{fig:tsp}
\end{figure}

We only consider symmetric TSPs where $d_{ij} = d_{ji}$. We associate nodes in our network with directed edges in the TSP tour. An edge from city $i$ to city $j$ is represented by node $(i,j)$. Each edge node has $3$ possible internal states, 3 input ports and 1 output port and an event from edge node $i,j$ means that this edge is present in the tour. Edge nodes are described by the state update function $f_{edge}$ and spike routing function $g_{edge}$ (see Fig.~\ref{fig:variable_b}) given by:
\begin{equation*}
  f_{edge}(i,s) =
  \begin{cases}
    3 & \text{if}\; (i=4 \; or \; (i=2 \; and \; s=2 )) \\
    2 & \text{if}\; i=3 \\
    1 & \text{if}\; (i=1 \; or \; (i=0 \; and \; s=3) \\
    s & \text{otherwise} 
  \end{cases}
  \quad\quad
  g_{edge}(i,s) =
  \begin{cases}
    1 & \text{if}\; (i=0 \; and \; s=3) \\
    0       & \text{otherwise} 
  \end{cases}
\end{equation*}
The node will only generate an event when it is in state $3$ (activated state) and when the internal oscillator generates an event. It then goes to state $1$ (deactivated state). The node will go to state $3$ if it receives an event on input port $4$ or if it is in state $2$ and receives an event on input port $2$.
We first consider a six cities symmetric TSP which can be mapped to the network shown in Fig.~\ref{fig:tsp_a}. Row $i$ contains edge nodes for edges originating from city $i$ and column $j$ edge nodes for edges that terminate at city $j+1$. Events from edge node $(i,j)$ are routed to port $1$ of all edge nodes in the same row and column, and to port $2$ on all edge nodes in row $j$. Assume initially all edge nodes are in state $2$ except the nodes in row $1$ which are in state $3$ (activated). One of the nodes in row $1$, for example $(1,3)$, will generate an event first and deactivate all edges nodes originating from city $1$ or terminating on city $3$ by putting these edge nodes in state $1$. The $1,3$ event activates all edge nodes originating from city $3$ (edge nodes in row $3$) by putting them in state $3$ (Fig.~\ref{fig:tsp_a}). Assume $(3,2)$ generates an event first in row $3$ thereby disabling all edges originating from city $3$ or terminating on city $2$. The event from $(3,2$) switches edge nodes in row $2$ that were in state $2$ to state $3$ (Fig.~\ref{fig:tsp_b}). One of the activated edge nodes in row $2$ will now generate an event and the sequence continues (Fig.~\ref{fig:tsp_c}). Through the events of the edge nodes (exactly $5$ nodes will generate an event), we obtain a valid tour, for example: $(1,3)$; $(3,2)$; $(2,4)$; $(4,6)$; $(6,5)$.  After these $5$ events, all edge nodes are in state $1$ (deactivated). 


The chance for an edge node $(i,j)$ that is at state $3$ to ``win the race'' and send an event first when row $i$ is activated depends on its frequency $f_{i,j}$  and the frequencies of the other active nodes in the row. As $f_{i,j}$ increases, the likelihood for edge node $(i,j)$ to generate an event first and become part of the tour when its row is activated increases. To increase the ``probability'' of obtaining a short tour, we set $f_{i,j}$ to:
\begin{equation*}
f_{i,j} = \frac{K}{d_{i,j}} + \eta_{i,j}
\end{equation*}
where $K$ is a positive scaling constants and $\eta_{i,j}$ is a small perturbation to keep the frequencies incommensurable. The node of a shorter edge has a higher frequency which makes the edge more likely to appear in the tour. 

The `tour completion' node in Fig.~\ref{fig:tsp_a} is responsible for resetting the network at the end of a valid tour. It has 2 internal states, $1$ input port, and $1$ output port and is defined as: 
\begin{equation*}
  f_{tc}(i,s) =
  \begin{cases}
    1 & \text{if}\; i=1 \\
    2 & \text{if}\; i=0
  \end{cases}
  \quad\quad
  g_{tc}(i,s) =
  \begin{cases}
    1 & \text{if}\; (i=0 \; and \; s=2) \\
    0 & \text{otherwise} 
  \end{cases}
\end{equation*}
The frequency of the `tour completion' node is chosen to be slightly less than that of the edge node having the smallest frequency and it receives events from all edge nodes. If the `tour completion' node does not receive any event within one oscillation cycle (all edge nodes are inactive), it generates an event that goes to port $4$ of the edge nodes in row $1$ and port $3$ of all other edge nodes thereby resetting the network and starting the tour generation process. 

Figure~\ref{fig:tsp_d} shows the frequency of occurrence of the tours generated by a network encoding a six city TSP. There are $120$ tours as the first city in the tour is always city $1$. Tours having the same distance do not occur equally often and some tours with longer distances occur more often than tours with shorter distances. In general, however, shorter tours tend to occur more often. 

For large problems, the performance of the TSP network falls far short of state of the art TSP algorithms. TSPs have a global constraint which is the requirement that a tour is valid. The architecture we describe is ill-suited to handling such a global constraint as different parts of the network can no longer operate in parallel to optimize the local constraints. As the size of the TSP increases, the fraction of valid tours decreases exponentially which makes it imperative that this global constraint be strictly enforced, otherwise the obtained tours will mostly be invalid. We followed a pseudo-sequential scheme, which is counter to the distributed nature of the architecture, where edges are added one by one to keep the global tour validity constraint satisfied at all times. Even though the TSP architecture we describe is inefficient, it shows that analog costs can be implemented through the appropriate choice of node frequencies and that the deterministic network can exhibit a sampling-like behavior. In implementations where individual node frequencies are not controllable, the same effect can be achieved by choosing nodes having the appropriate frequency relations from a large pool of mismatched nodes. 

\subsection*{Prototype VLSI implementation}
The prototype VLSI chip that implements a version of the architecture described in this paper is composed of a 2D array of binary nodes that communicate using events. The problem of transmission and routing of asynchronous events has been thoroughly investigated in the neuromorphic engineering literature~\cite{Deiss_etal98,Boahen00}. An elegant solution uses a communication protocol based on the Address-Event Representation (AER). 
When a node generates an event on one of its output ports, it executes a handshake protocol with the `output AER interface'. The `output AER interface' encodes the address of the output port on which the event was generated and transmits the address off-chip using an output bus that has $log_2(K_{out})$ lines. $K_{out}$ is the number of possible event sources (the output ports of all the nodes). The array has $K_{in}$ possible event targets (the input ports of the nodes), if an event is to be sent to one of these targets, the target address is sent to the `input AER interface' on a bus that has $log_2(K_{in})$ lines. The `input AER interface' decodes the address and sends an event to the target element by simultaneously activating the correct row and column in the array.

The 2D array on the chip comprises 64*32 binary nodes/variables, i.e, nodes/variables with two output ports. The chip can be configured so that 2,3, or 4 adjacent variables are merged together to realize 4-, 6-, or 8-valued variables respectively. An n-valued variable ($n \in \{2,4,6,8\}$) has $n$ output ports and $n$ possible internal states and has $2^n-1$ input ports. Physically, a variable has $n$ digital input lines on which it receives a binary word encoding the index of the input port receiving the event. 
An off-chip event router implemented on a field programmable gate array (FPGA) communicates with the output and input AER interfaces to route events from nodes/variables output ports to input ports according to a programmable routing table. 

When an n-valued node/variable receives an event on port $i$, The $1$s in the binary representation of $i$ denote the allowable internal states that the variable can take. The variable has $n$ possible internal states and an event on one of the  $2^n-1$ input ports can thus decide which non-empty subset of these states are allowed. If multiple states are allowed, the variable stays at its current state if the current state is one of the allowed states, otherwise it goes to the lowest index allowed state. Let $i(p)$ be the $p^{th}$ bit of $i$ where indexing starts at 1, the state update function $f$ is thus:
\begin{equation}
  f_{HW}(i,s) =
  \begin{cases}
    s & \text{if}\;\;\; i(s)=1\;\; or \;\; i=0 \\
    p       & \text{if}\;\;\; (i(s)=0\;\; and\;\; i(p)=1\;\; and \;\; i(j)=0) \;\;\text{for}\;\;\ j=1,2,\ldots,p-1 \\
  \end{cases}
 \label{eq:hwstate}
\end{equation}
The node/variable generates an event only when it receives an event on port $0$. The event is generated on the port corresponding to the currents state. The event routing function $g$ is:
\begin{equation}
  g_{HW}(i,s) =
  \begin{cases}
    s & \text{if}\; i=0 \\
    0       & \text{otherwise} 
  \end{cases}
 \label{eq:hwroute}
\end{equation}

\begin{figure}[h]
 \centering

 \begin{subfigure}[b]{0.3\textwidth}
      \includegraphics[width=\textwidth]{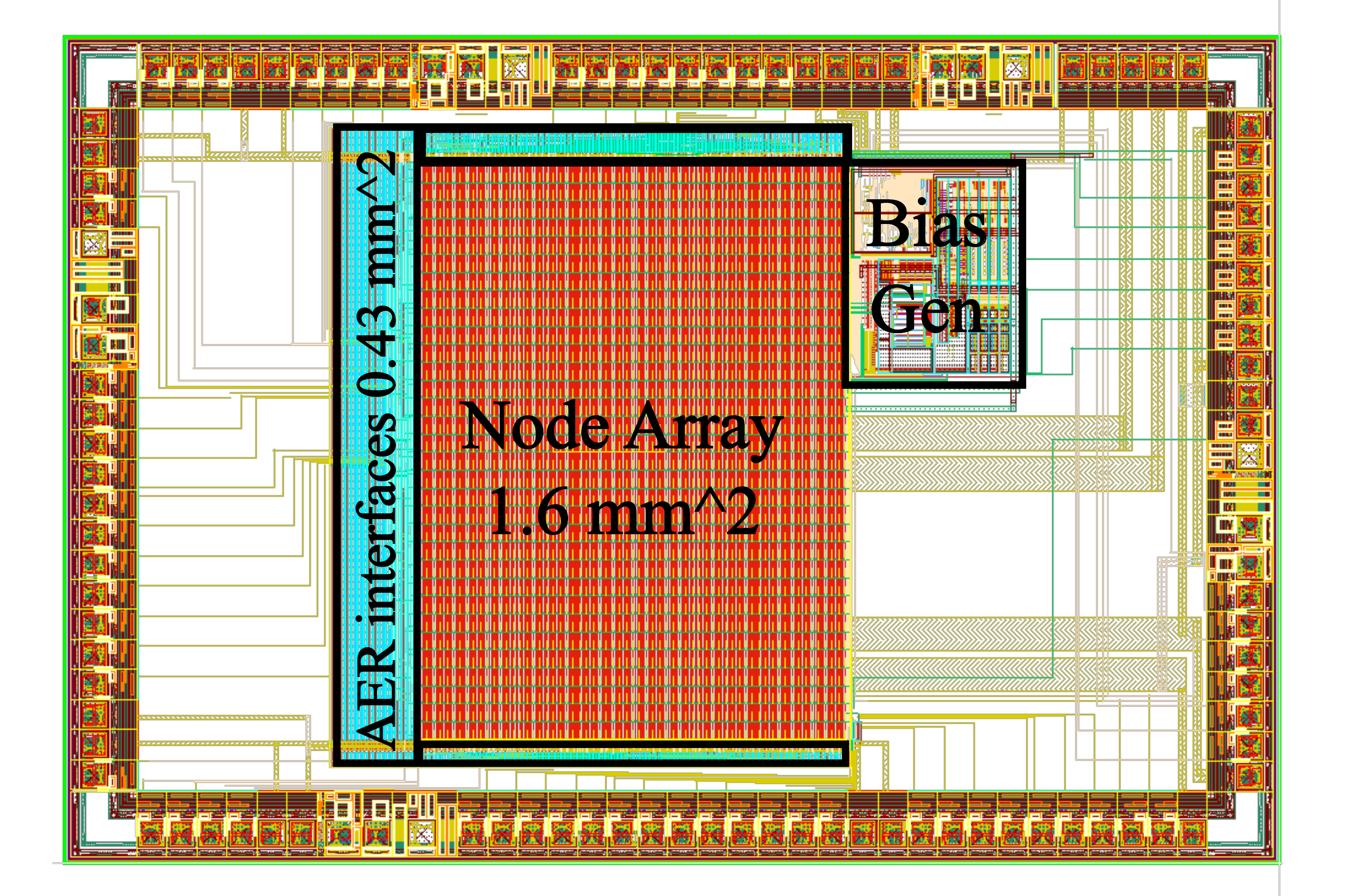} 
      \subcaption{}
      \label{fig:arch_a}
    \end{subfigure}
    \quad
 \begin{subfigure}[b]{0.3\textwidth}
     \includegraphics[width=\textwidth]{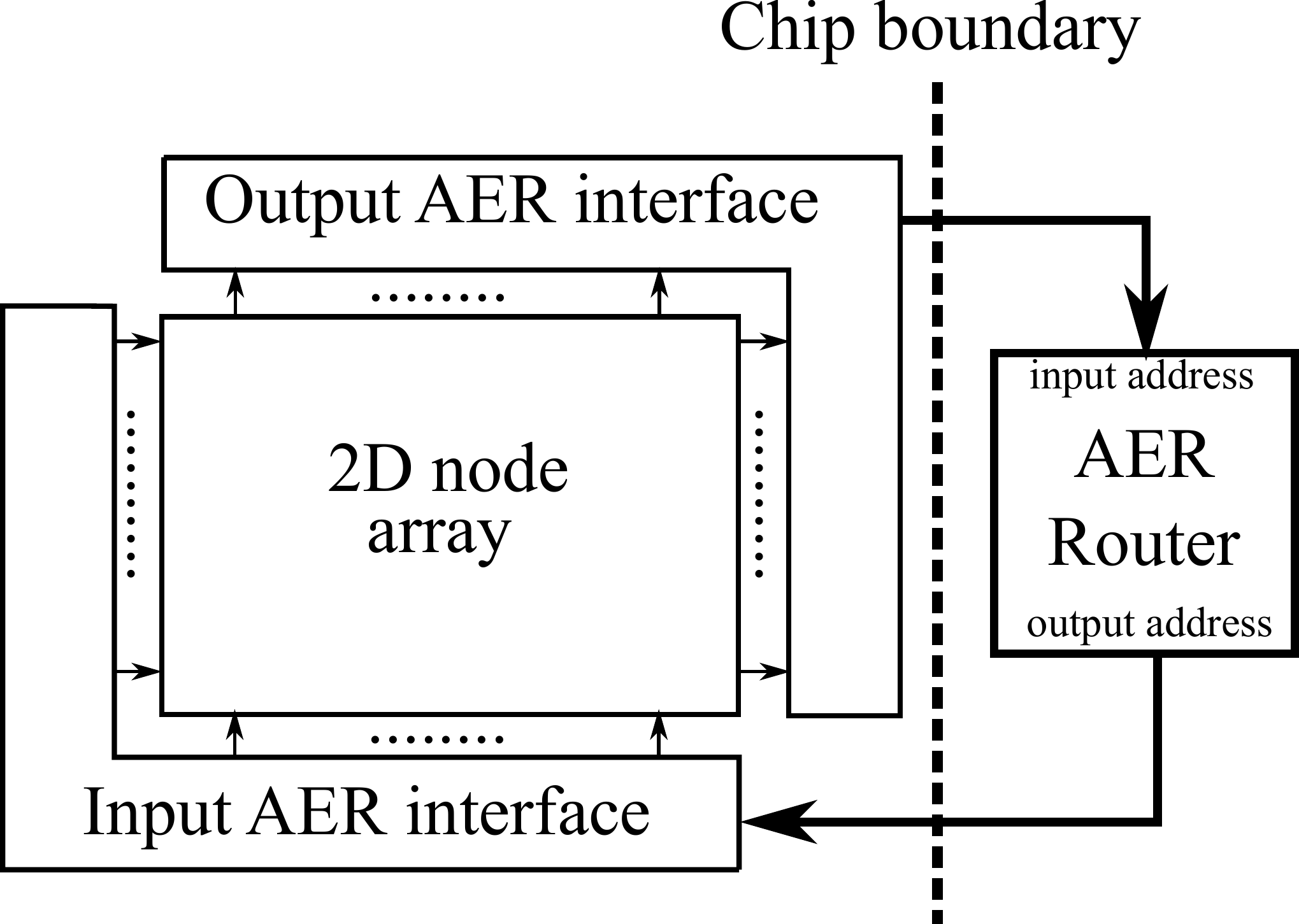} 
     \subcaption{}
     \label{fig:arch_b}
   \end{subfigure}
   \quad
   \begin{subfigure}[b]{0.3\textwidth}
     \includegraphics[width=\textwidth]{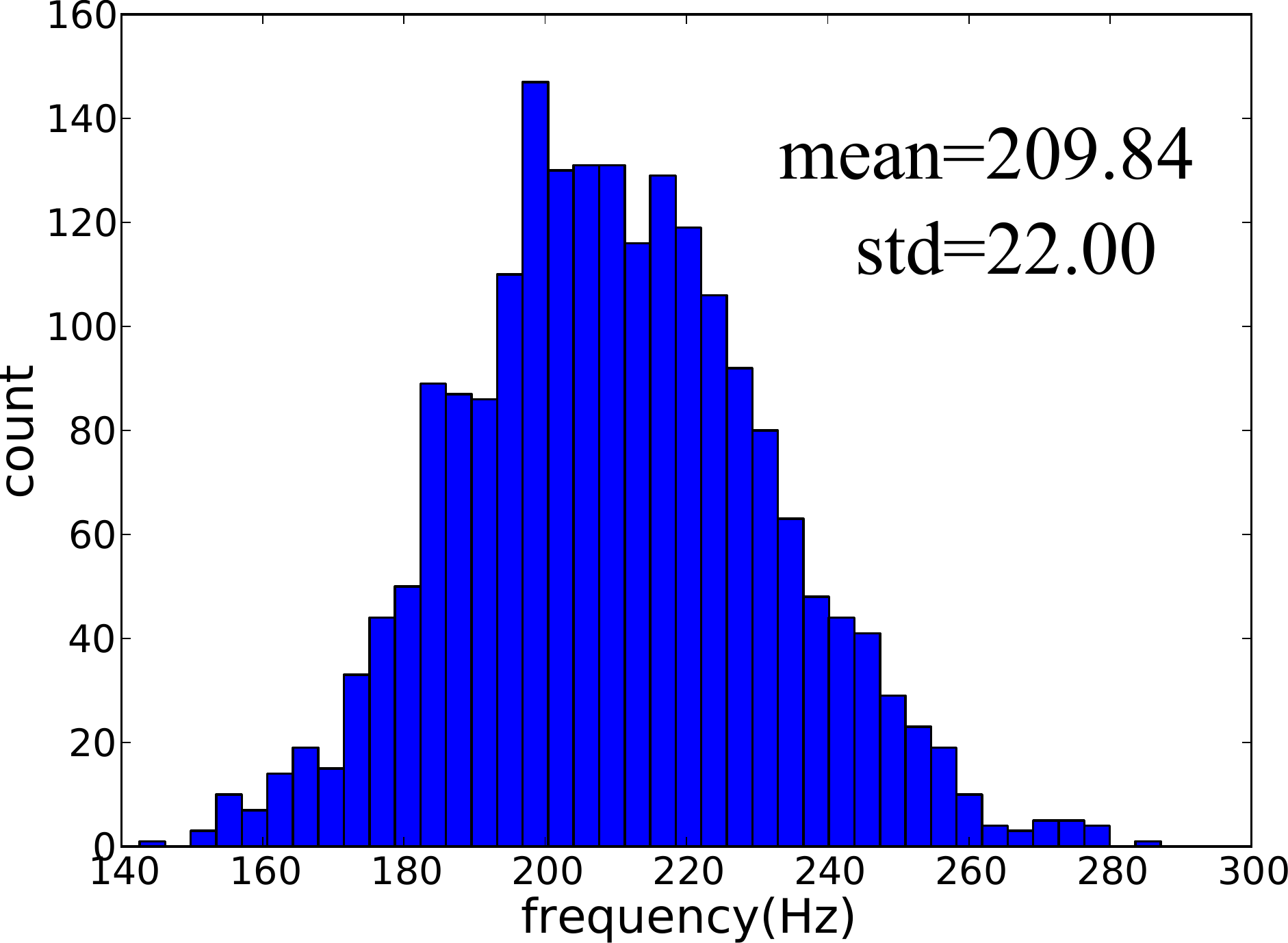} 
     \subcaption{}
     \label{fig:arch_c}
   \end{subfigure}

   \caption{(\subref{fig:arch_a}) Layout of the minimum size (2mm*3mm) prototype chip fabricated using a $180$\, nm CMOS process that implements the architecture described in this paper. The 64*32 node array in the middle is surrounded on three sides by the digital asynchronous AER interfaces. An externally programmable bias generation block generates the analog biases needed by the analog oscillators.(\subref{fig:arch_b}) An off-chip event router implemented on an FPGA communicates with the chip AER interfaces to route events from output ports to input ports. (\subref{fig:arch_c}) Frequency distribution of the 2048 analog oscillators on-chip for the bias conditions used in the experiments in this paper.}
\label{fig:arch}
\end{figure}

The analog oscillator in each variable is realized using an integrate and fire neuron~\cite{Indiveri_etal06} receiving constant current injection. As shown in Fig.~\ref{fig:arch_c}, the oscillator frequencies are significantly different due to transistor mismatch. Since the oscillation frequencies are real numbers drawn from a probability distribution arising from the variability inherent in the fabrication process, it is impossible for an oscillator to have a frequency that is a rational multiple of another's.

\subsubsection*{Solving 3-SAT problems on the hardware prototype}
Using the node logic on the prototype chip, we implemented a simple 3-SAT algorithm. Each problem variable is represented by one binary node and each 3-SAT constraint/clause is represented by a 4-valued node. An event from the $1$ port or the $2$ port of a binary variable/node denotes that the variable value is $0$ or $1$ respectively. The constraint/clause node is in state $4$ if the constraint is fulfilled, otherwise its state ($1$ or $2$ or $3$) denotes which literal in the constraint last emitted an event. Consider a 3-SAT constraint $C1 = (L1 \vee L2 \vee \neg L3)$. Events from port $2$ of variables/nodes $L1$ and $L2$ and events from port $1$ of $L3$ should put the $C1$ node at state $4$ (constraint fulfilled). A complementary event, i.e, an event that does not cause the constraint to be fulfilled (for example a $2$ event from $L3$) should do nothing if the constraint is fulfilled as we assume one, or both, of the other two variables fulfill the constraint. However, if the constraint is not fulfilled, a complementary event from the $k^{th}$ variable in the constraint should put the constraint node in state $k$. 

When the constraint node advertises its state by an event, events from ports $1$, $2$, or $3$ should set the first, second, or third variable respectively to a constraint-fulfilling state. In the scheme described so far, when a constraint node is fulfilled (in state $4$), events from the variables will never move it away from the fulfilled state. To address this, whenever the constraint node generates an event on port $4$, this event is routed back to the constraint node and moves it to an arbitrary unfulfilled state (we arbitrarily choose state $3$). Thus, within each oscillation cycle of the constraint node, the node has to receive a constraint-fulfilling event from one of its variables in order to go to state $4$ and in order not to generate an event at the end of the oscillation cycle that forces one of these variables to fulfill the constraint. The constraint nodes were picked from among the nodes with the lowest oscillation frequencies. The globally optimal solution is thus stable as the variable(s) fulfilling a constraint will always be able to generate at lease one event that puts the constraint node in a fulfilled state during each cycle of the constraint node.

\begin{figure}[h]
 \centering
 \begin{subfigure}[b]{0.4\textwidth}
     \includegraphics[width=\textwidth]{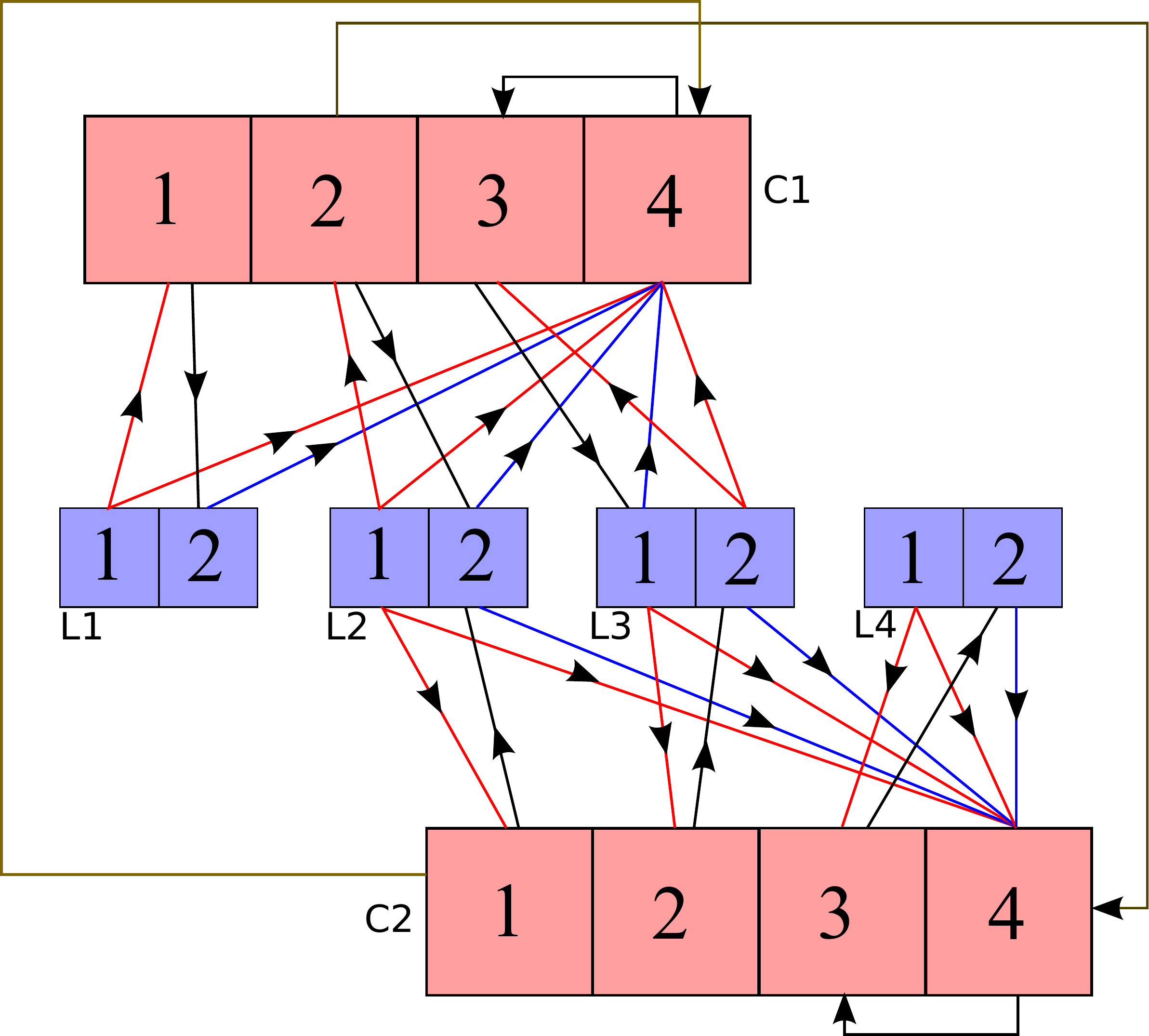} 
     \subcaption{}
     \label{fig:hwsat_a}
   \end{subfigure}
   \quad
   \begin{subfigure}[b]{0.5\textwidth}
     \includegraphics[width=\textwidth]{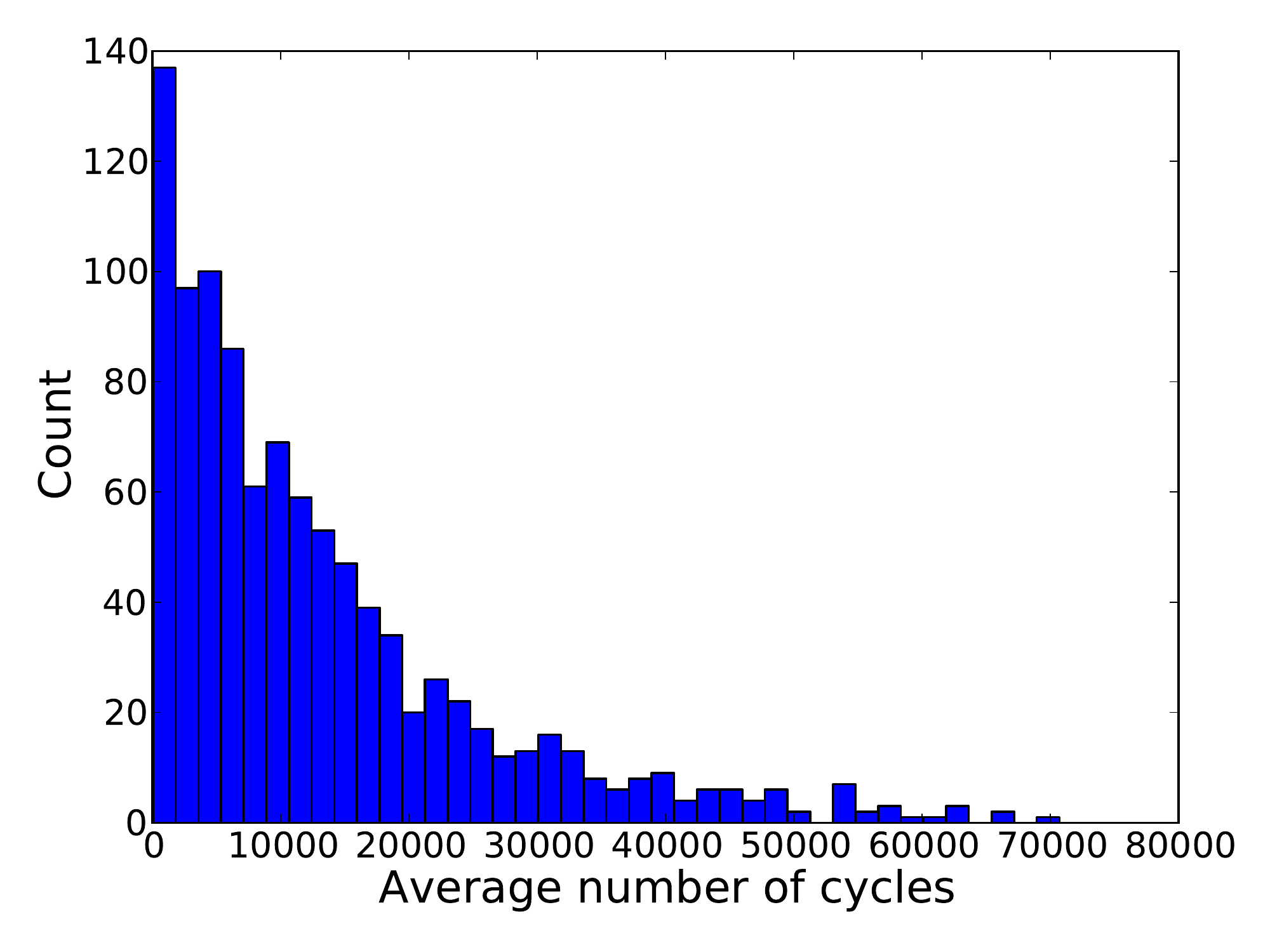} 
     \subcaption{}
     \label{fig:hwsat_b}
   \end{subfigure}

   \caption{(\subref{fig:hwsat_a}) Network showing the implementation of $C1 \wedge C2$ where $C1 = (L1 \vee L2 \vee \neg L3)$ and $C2 = (L2 \vee L3 \vee L4)$. Numbered squares indicate the output ports and, indirectly, the input ports of a variable and arrows indicate how events are routed. For example, events from port $1$ of $L1$ go to input port $9$( '1001' in binary) of $C1$ which instructs $C1$ to go to state $4$ or state $1$. Events from port $2$ of $L1$ go to port $8$ of $C1$ which instructs $C1$ to go to state $4$. (\subref{fig:hwsat_a}) Histogram of the number of oscillation cycles (averaged per variable) needed by the chip to find the solution of a 3-SAT problem with 50 variables and 218 clauses over 1000 trials taken from \cite{Hoos_Stutzle98}}
\label{fig:hwsat}
\end{figure}

The above scheme is implemented by routing events according to Fig.~\ref{fig:hwsat_a}. Events from variable nodes can not dislodge a constraint node from the fulfilled state or state $4$. Note that state $4$ of a constraint node is the lowest priority state according to the state update function in Eq.~\ref{eq:hwstate} so an input event to a constraint node which encodes that state $4$ and state $k$ ($k \in \{1,2,3\}$) are allowed will always put the constraint node in state $k$, if it was not already at state $4$. If a variable appears with the same sign in multiple constraints (negated or non-negated in all of them), an event generated by one of these constraint nodes that forces this common variable to go to a fulfilling state will automatically fulfill the other constraints as well so we route such events to the other constraints so as to move them to the fulfilled state as shown in Fig.~\ref{fig:hwsat} and prevent them from unnecessarily flipping other variables. This scheme for solving 3-SAT is less powerful than the previously described approach based on the probSAT algorithm. At the end of the cycle of an unfulfilled constraint node, the constraint node simply flips the last variable in its domain to generate an event. Due to the continuously shifting phase relations, the choice of which variable to flip is essentially done ``at random''  with no regard for how many other constraints would be violated due to this flip. Figure~\ref{fig:hwsat_b} shows a histogram of the average number of oscillation cycles per variable needed to find the solution of an example 3-SAT problem.

\subsubsection*{Solving graph coloring problems on the hardware prototype}
\begin{figure}[h]
 \centering
 \begin{subfigure}[b]{0.4\textwidth}
     \includegraphics[width=\textwidth]{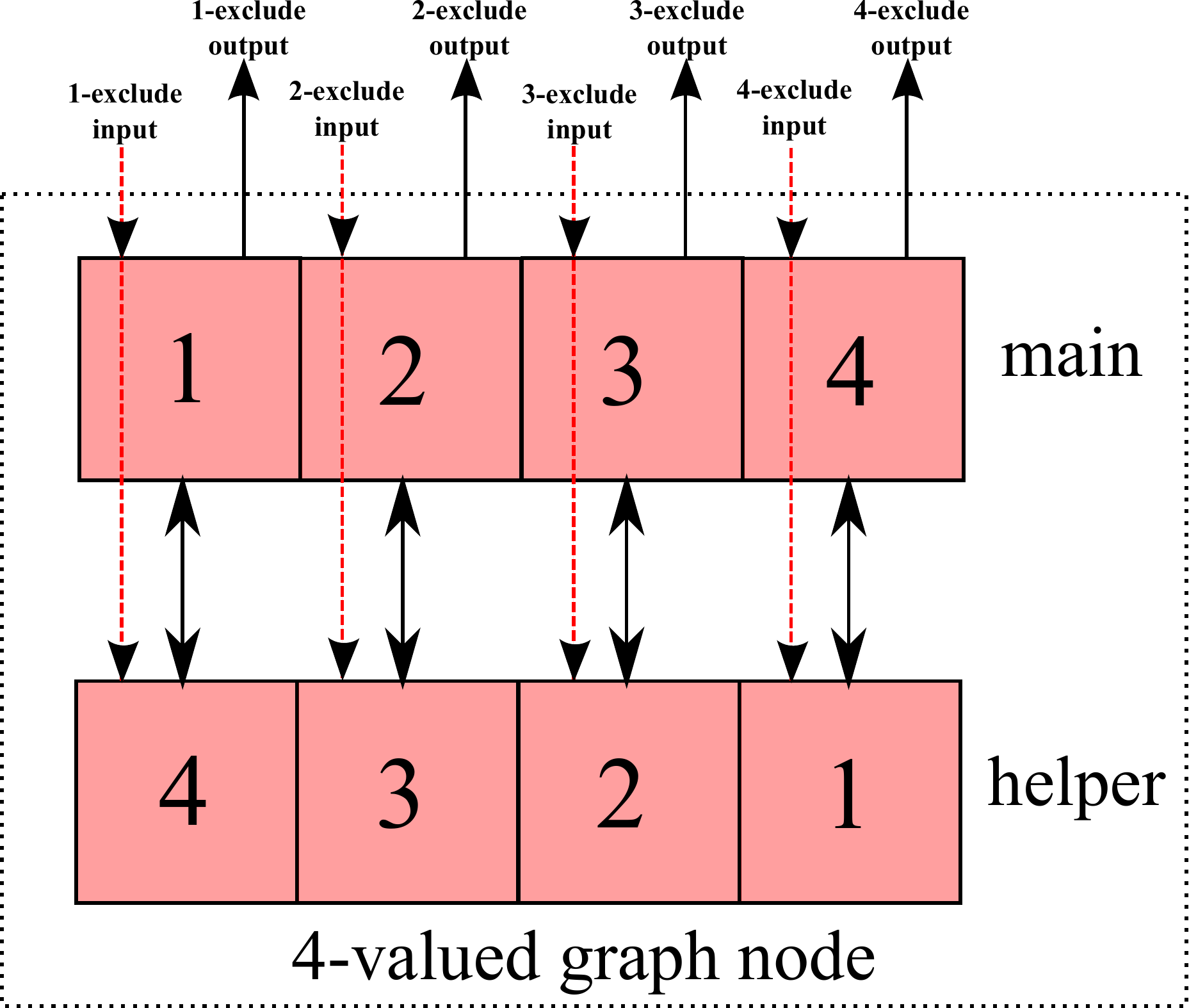} 
     \subcaption{}
     \label{fig:hwcoloring_a}
   \end{subfigure}
   \quad
   \begin{subfigure}[b]{0.4\textwidth}
     \includegraphics[width=\textwidth]{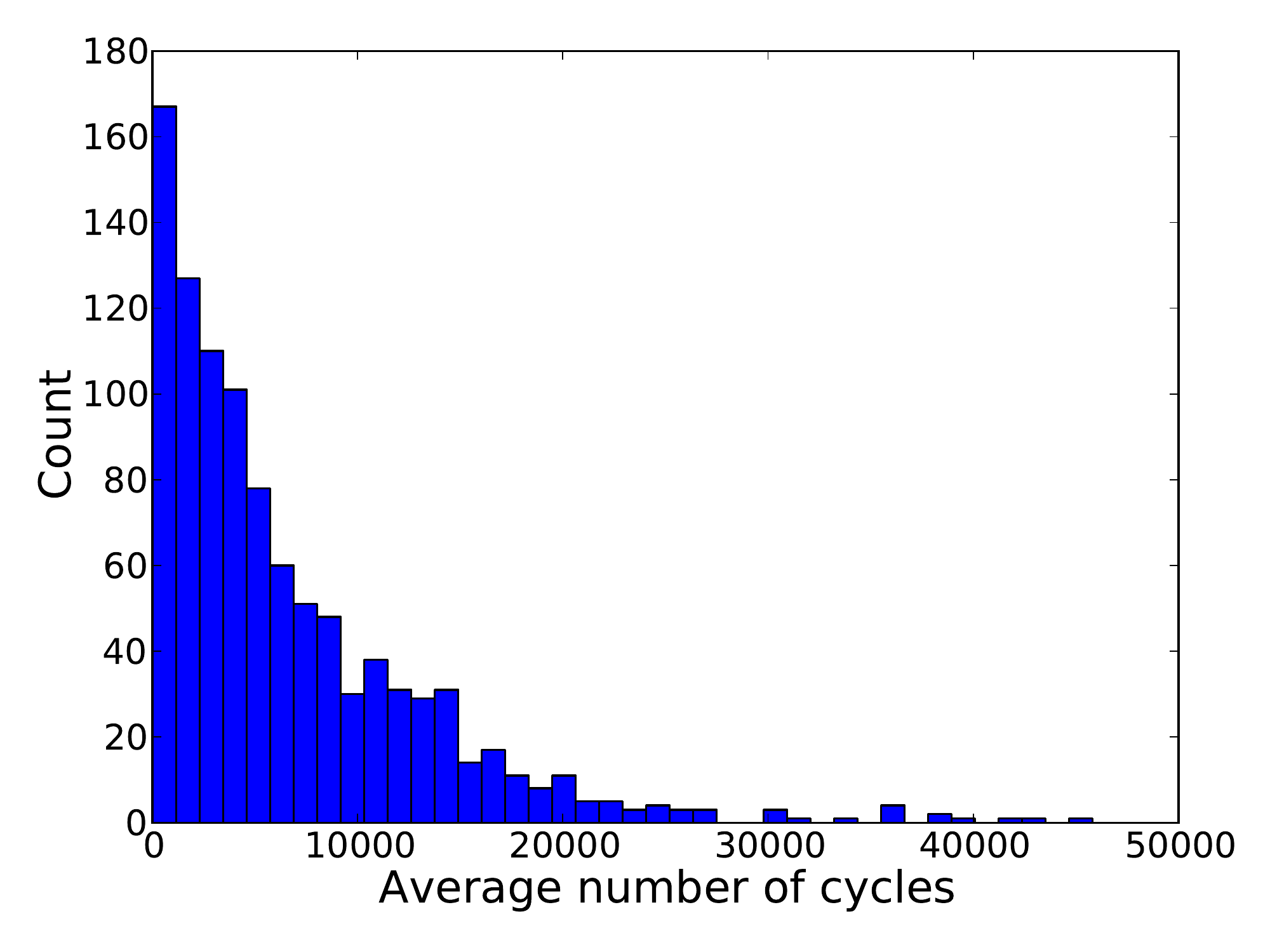} 
     \subcaption{}
     \label{fig:hwcoloring_b}
   \end{subfigure}

   \caption{(\subref{fig:hwsat_a}) Implementation of a 4-valued graph vertex using two 4-valued chip nodes which are coupled so that an event from port $1$, $2$, $3$, or $4$ of one chip node puts the other node in state $4$, $3$, $2$, or $1$ respectively. This vertex receives events from other vertices which go the exclude input ports of the two chip nodes (red dashed lines). For example an event arriving on the 1-exclude input port goes to port 14(binary '1110') on the `main' chip node and port 7(binary '0111') on the `helper' chip node. (\subref{fig:hwsat_a}) Histogram of the number of oscillation cycles (averaged per variable) needed by the chip to find the optimal coloring of the $5\times 5$ queen graph.}
\label{fig:hwcoloring}
\end{figure}

The hardware prototype can solve graph coloring problems with up to 8 colors. We can use a simple scheme where a graph vertex is represented by a chip node and events from port $p$ of a node (which indicates that the node is in state/color $p$) go to input port $2^n-1-2^{p-1}$ (all $1$s binary string except at position $p$; $p$ index starts from 1) of all adjacent nodes/vertices in the graph. We call these input ports the i-exclude input ports as receiving an event on them instructs the node to go to any state except $p$, thereby enforcing the constraint. However, this scheme will not work since a node always goes to an allowed state that has the lowest index when responding to an exclude event (see Eq.~\ref{eq:hwstate}). All the nodes would thus quickly get stuck in the $1$ and $2$ states as the 1-exclude and 2-exclude events that the nodes send to each other will not be able to move any node out of these 2 states. We use the more elaborate scheme shown in Fig.~\ref{fig:hwcoloring_a} where two 4-valued chip nodes are used to implement one 4-valued graph vertex. The value of this graph vertex is index of the last event emitted by the `main' chip node. 

Pair-wise inequality constraints are implemented by routing events from the i-exclude output port of one vertex to the i-exclude input port of the other vertex. Assume a vertex has value $1$, i.e, the state of the main (helper) chip nodes are $1$ ($4$). The state/color of this vertex will only change if it receives an event on the 1-exclude port. In that case, the `main' and `helper' chip nodes go to states $2$ and $1$ respectively  since these are the lowest index allowed states in the two chip nodes. The two chip nodes now have inconsistent states and whichever of them generates an event first forces the other node to switch its state; for example if the `helper' node generates an event first, it forces the `main' node to take state $4$. A 1-exclude input event effectively has a $50\%$ chance of moving this graph node to state $2$ and a $50\%$ chance to move it to state $4$ due to the irregular phase relations.

The scheme can be extended to 6- and 8-valued vertices by using three 6-valued and four 8-valued chip nodes respectively to represent a single graph vertex and it is straightforward to show that using this scheme, the network representing the coloring graph always uses all available colors. 3-, 5-, and 7-coloring problems can be implemented by adjusting the even color schemes so that events are routed to input ports that exclude both the color/index of the source output port as well as the highest index/color which will then be unused. 

One difficult graph for this architecture is the `$5\times 5$ queen' graph whose solution is equivalent to finding the non-interfering positions of 5 queens on a $5 \times 5$ chess board. The average number of cycles needed to find a solution is shown in Fig.~\ref{fig:hwcoloring_b}.

\subsection*{Discussion and Conclusions}
\label{sec:discussion}

CSPs have often been examined through the lens of statistical physics~\cite{mezard_etal02,krzakala_kurchan07}. Within the framework of statistical physics, a CSP is formulated as a distributed system that seeks to minimize the number of frustrated interactions (violated constraints) between its elements. Direct analogies can be established between the ground energy states of physical systems (where frustrated interactions are at a minimum) and solutions to CSPs~\cite{Barahona82}. The architecture we describe in this paper is fundamentally different from the systems analyzed in the framework of statistical physics, yet it captures some of the general features of such systems: The architecture makes use of a large number of locally interacting elements that mutually constrain each other so that the system as a whole tries to go to states where the number of frustrated interactions is at a minimum. 

Perhaps the most distinguishing feature of our system is the mechanism used to explore the solution space. In lieu of random fluctuations, the continuously varying phase relations between incommensurable oscillators are a source of non-repeating fluctuations that can be easily exploited in our event based architecture to realize efficient search algorithms.  True- or pseudo-random number generators require significant hardware resources, in terms of power and silicon area. The paradigm described in this paper is more efficient, as only  mismatched oscillator circuits are needed. The transistor mismatch inherent in VLSI electronic circuits ensures that the fabricated oscillator circuit frequencies are incommensurable. While circuit designers typically struggle to minimize the effects of device mismatch by using larger devices, in our case mismatch is a beneficial property whose effects should be preserved, thereby simplifying the implementation of the oscillators/variables and minimizing their area on silicon.

The digital event-based nature of variable communication is key to the architecture's scalability and configurability. These digital pulses can be transmitted and routed using a digital fabric that links together a large number of nodes/variables. In the prototype chip, event routing is done off-chip in a serial manner on the FPGA. This introduces a serial bottleneck in the otherwise massively parallel operation of the architecture. However, configurable and parallel AER routing fabrics are already in use in large-scale neuromorphic systems~\cite{Joshi_etal10,Merolla_etal14} and could be directly adapted for use in an implementation of the described architecture. As shown when solving SAT problems, the architecture is robust to event delays and lost events which relaxes the requirements on the event routing fabric. 

In simulation we showed for the case of SAT problems that the proposed architecture can run at a surprisingly slow mean oscillation frequency (around 0.2-2 MHz) and still attain a time to solution that is comparable to a CPU running at three orders of magnitude higher clock rate. The simple logic operations in the constraint and literal nodes can certainly run at such slow frequencies. These results indicate that the proposed architecture is a far more efficient approach to solving SAT problems than conventional CPUs. 

Algorithms for solving CSPs are often conceived with the digital von Neumann model of computation in mind. The results presented in this paper highlight an alternative approach which starts with no prior assumptions about the computational model, and seeks to exploit the physical characteristics of the underlying substrate in order to find a solution tailored to the computational problem at hand. In our case, we exploited the natural incommensurability of physical analog oscillators to derive a distributed novel algorithm for solving CSPs. This algorithm naturally results in an efficient implementation in the physical substrate that underlied its derivation. The computing architectures developed using this bottom-up approach, such as the VLSI device we present in this paper, have the potential to achieve considerable performance gains in their target problems compared to conventional purely digital approaches.

\bibliographystyle{unsrt}

\section*{Acknowledgements}
\label{sec:acknowledgements}

This work was supported by the European CHIST-ERA program, via the ``Plasticity in NEUral Memristive Architectures'' (PNEUMA) project and by the European Research council, via the ``Neuromorphic Processors'' (neuroP) project, under ERC grant number 257219.

\end{document}